\documentclass[superscriptaddress,groupedaddress,nofootnoteinbib,12pt]{article}  

\usepackage{graphicx}  
\usepackage{dcolumn}   
\usepackage{bm}        
\usepackage{jcappub}
\usepackage[normalem]{ulem}

\def\be{\begin{equation}}
\def\ee{\end{equation}}
\def\ba{\begin{eqnarray}}
\def\ea{\end{eqnarray}}
\def\beq{\begin{eqnarray}}
\def\eeq{\end{eqnarray}}

\def\noi{{\noindent}}
\definecolor{verde}{rgb}{0,0.5,0}
\def\blu{\color{blue}}

\def\p{{\cal P}}

\def\L*{{\cal L}_*}
\def\L{\mathcal{L}}
\def\({\left(}
\def\){\right)}

\def\p{\partial}

\def\p{\partial}

\def\<{\langle}
\def\>{\rangle}

\def\cs2{c_{s}^{2}}

 \def\p{\partial}

 \def\be   {\begin{equation}}   \def\ee   {\end{equation}}
 \def\ba   {\begin{array}}      \def\ea   {\end{array}}
 \def\bea  {\begin{eqnarray}}   \def\eea  {\end{eqnarray}}
 \def\bean {\begin{eqnarray*}}  \def\eean {\end{eqnarray*}}

\usepackage{xcolor}
\definecolor{MatteoColour}{rgb}{0.,0.7,0.3}

\begin{document}

\title{Possible Signatures of Inflationary Particle Content: Spin-2 Fields}

\author{Matteo~Biagetti$\,^{a}$,}
\author{Emanuela~Dimastrogiovanni$\,^{b}$,}
\author{Matteo~Fasiello$\,^{c}$}
\affiliation{$^{a}\,$ 
Institute of Physics, Universiteit van Amsterdam, Science Park, 1098XH Amsterdam, The Netherlands}
\affiliation{$^{b}\,$CERCA \& Department of Physics, Case Western Reserve University, Cleveland, OH 44106, USA}
\affiliation{$^{c}\,$Stanford Institute for Theoretical Physics and Department of Physics, Stanford University, Stanford, CA 94306}



\abstract{We study the imprints of a massive spin-2 field on inflationary observables, and in particular on the breaking of consistency relations. In this setup, the minimal inflationary field content interacts with the massive spin-2 field through dRGT interactions, thus guaranteeing the absence of Boulware-Deser ghostly degrees of freedom. The unitarity requirement on spinning particles, known as Higuchi bound, plays a crucial role for the size of the observable signal.
\vspace{4cm}}

\maketitle


\section{Introduction}
\label{intro}
The main reason for investigating non-minimal particle content during inflation is the energy scale at which inflationary dynamics may take place. With high-scale inflation set at about $10^{14}{\rm \,GeV}$, there is no other comparable window on such high energy processes. The characterization of the signatures corresponding to different particle species is indispensable for a precise observation-to-theory mapping in case of detection. Cosmological probes can indeed inform our model building and even UV completion efforts. This line of reasoning is at the heart of recent literature on the subject \cite{Kehagias:2015jha,Arkani-Hamed:2015bza,Dimastrogiovanni:2015pla,Lee:2016vti,Baumann:2014nda}. 

Additional degrees of freedom (dof) during inflation may take many shapes (mass, spin). Perhaps the most immediate indication of their impact on observables stems from the value of their mass as compared to the Hubble scale $H$. The latter is indeed a crucial, threshold, value. The reasons are manifold. First of all, the energy scale of inflation is itself generally of order $H$; it naturally follows that too large a mass, $m\gg H$, would justify the integrating out of any dof leaving very little in terms of imprints (although see e.g. \cite{Achucarro:2012sm,Burgess:2012dz,Silverstein:2017zfk} for important caveats). A mass of order Hubble can instead support distinct signatures and has been the subject of many studies, especially in the scalar case \cite{Chen:2009we,Chen:2009zp,Craig:2014rta,Noumi:2012vr,Sefusatti:2012ye,McAllister:2012am,Dimastrogiovanni:2015pla,Baumann:2011nk}. This range of values for the mass is also natural from the point of view of supersymmetric theories: if such symmetry is not broken at a higher scale, the inflationary vacuum energy will eventually break it, thereby justifying the expectation that some of the fields populating the supersymmetric multiplets will have masses of order Hubble. The behaviour of inflationary correlation functions also dictates that particles whose Compton wavelength is much smaller than the horizon \footnote{The equations of motion, for example in quasi-de Sitter, show that as soon as $m\sim H$ the mass of a generic particle becomes relevant to its dynamics and leads to decaying solutions.} have negligible effects on late-time observables, identifying again $1/H$ as a lower limit.  
Finally, for particles of spin $s=2$ or higher, unitarity imposes lower bounds on the mass in de Sitter 
\bea
m^2> s(s-1) H^2,
\eea
which further underscores the frontier nature of the $m\sim H$ mass range. The unitarity or Higuchi \cite{Higuchi:1986py,Blas:2007zza,Comelli:2012db} constraint is less stringent away from de Sitter \cite{Fasiello:2012rw,Fasiello:2013woa} but, as we shall see, it remains a powerful bound.

The most efficient way to probe both the field content and inner structure of the inflationary Lagrangian is to consider N-point correlation functions, non-Gaussianities ($N\geq 3$). The Planck satellite mission has set precise bounds on non-Gaussianities, starting with the three-point function of curvature fluctuations \cite{Ade:2015ava} in the key momenta configurations such as local, equilateral, orthogonal. Within observationally viable values, there is still plenty of room to seek imprints of dynamics beyond the single-field inflation scenario. Of particular interest are the squeezed limit of the three-point function and its higher order generalizations. This configuration is particularly sensitive to the presence of additional dof in light of consistency relations (CRs). The latter are relations between the squeezed limit of an $N+1$-point function and its N-point counterpart.

 CRs stem from residual gauge diffeomorphisms of the (inflationary) theory \cite{Maldacena:2002vr,Creminelli:2004yq,Assassi:2012zq,Hinterbichler:2013dpa,Tanaka:2017nff}.  One may derive non-trivial CRs when the soft mode characterizing any squeezed limit transforms non-linearly under
the residual diff. In the minimal inflationary setup
the curvature perturbation $\zeta$ and tensor mode $\gamma$ transform non linearly respectively under dilatation and anisotropic rescaling, thereby generating the well-known scalar and tensor CRs. In each of these two cases, the leading effect of a long scalar (tensor) mode on $N$ short modes corresponds to the action of a residual gauge-symmetry on the $N$-point function and can therefore be gauged away.\\
In this work we focus on the case of a three-point function with a long tensor mode and two scalar short modes. In this case, the CR in single-field inflation reads 
\begin{equation}\label{tensorCR}
\lim_{\vec k_1\rightarrow 0} \Big< \gamma^\lambda_{\vec k_1}\, \zeta_{\vec k_2}\,\zeta_{\vec k_3}\Big>'\simeq \frac 32  \Big< \gamma^\lambda_{\vec k_1}\, \gamma^\lambda_{-\vec k_1}\Big>'\, \epsilon^\lambda_{ij}\, \hat k_{3i}\,\hat k_{3j}\, \Big<  \zeta_{\vec k_2}\,\zeta_{\vec k_3}\Big>',
\end{equation}
where $\gamma$ and $\zeta$ are the tensor and scalar perturbations respectively,  we defined the polarization tensors $\epsilon^\lambda_{ij}$ as $\gamma_{ij} = \sum_\lambda \epsilon^\lambda_{ij}\,\gamma^\lambda$ and the prime indicates that we left implicit the momentum-conserving delta.\\
\indent Whenever additional fields appear on the scene, CRs can be broken\footnote{For our purposes it suffice to have non-standard consistency relations in the sense that the variation of the power spectrum is not related to the squeezed limit of one three-point function but rather to a weighted linear combinations of at least two such observables.} if, for example, there exists one more independent mode that transforms non-linearly  under the diffeomorphism. It follows that the squeezed bispectrum may no longer be gauged away and an enhanced signal is made possible in this configuration. The power of CRs breaking as a probe of new physics is not limited to inflationary setups and has been widely employed in other contexts such as large scale structure \cite{Kehagias:2013yd,Peloso:2013zw,Kehagias:2013rpa,Creminelli:2013mca,Kehagias:2013paa,Creminelli:2013nua,Horn:2014rta,Kehagias:2015tda,DiDio:2016gpd,Fasiello:2016yvr}. One should further note that the squeezed one is the best  suited of configurations to reconstruct the inflationary potential as it is less sensitive to possible variations of the statistics of the fluctuations in sub-volumes \cite{Bonga:2015urq}.

In this manuscript we will report on work where the inflationary dynamics is enriched by a massive spin-2 field non-minimally coupled to the usual massless graviton via ghost-free interactions \cite{deRham:2010ik,deRham:2010kj,Hassan:2011tf}. As a result, the standard tensor CR is modified. We quantify the consequences of this ``breaking" on the squeezed limit of the tensor-scalar-scalar bispectrum signal and show how the constraints originating from a (softened) Higuchi bound propagate all the way to observables. Despite a weakened unitarity constraint  away from dS, we find that the amplitude of the signal produced in this setup is too weak for detection by upcoming large scale structure surveys.\\

This work is organized as follows. In \textit{Section \ref{model0}} we introduce the model under study comprising the inflaton field minimally coupled to a spin-2 field interacting with another (massive) spin-2 field through dRGT-interactions. We discuss at length the features of the model that will be most relevant in the inflationary context. \textit{Section \ref{calc}} is devoted to outlining how consistency relations play out in our specific setup and calculating the resulting tensor-scalar-scalar bispectrum. In \textit{Section\,\ref{obs}} we discuss the observational consequences of these results. We conclude with \textit{Section\,\ref{conclusions}}, where we comment on our findings and suggest possible directions for further work on the subject. In \textit{Appendix \ref{crs}} we provide a general overview of CRs and their breaking. In \textit{Appendix \ref{route1}} we detail on a route to modified CRs which is alternative, up to horizon scales, to the one in \textit{Section \ref{calc}}, and include a diagrammatic proof of CRs breaking.

\section{The framework}
\label{model0}
It is well-known that additional degrees of freedom populating the inflationary Lagrangian have the potential to modify CRs in favour of a enhanced squeezed bispectrum signal (see \textit{Appendix \ref{crs}} for a general discussion). There already exist several studies on the subject in the literature. We focus our analysis here on the case where the minimal inflationary particle content is enriched by a spin-2 field. Our discussion ought to start from the realization due to \cite{Boulanger:2000rq} that a theory of interacting\,\footnote{Note that there is a further implicit restriction: even  indirect, e.g. mediated by other fields, interactions are forbidden.} spin-2 particles admits at most one massless spin-2 field. As we are working from the perspective of adding particle content to the minimal (GR + scalar field) inflationary scenario, we are then immediately led to introducing a new, massive, graviton. 

The theory of massive gravity has received enormous attention in recent years. This is due to the combination of experimental results \cite{Riess:1998cb,Schmidt:1998ys,Perlmutter:1998hx} confirming the current acceleration of the universe and the recent non-linear formulation \cite{deRham:2010ik,deRham:2010kj} of a ghost-free massive gravity theory known as dRGT. An infrared modification of gravity is indeed among the most studied mechanisms, along with dark energy, to explain late time acceleration. The interest for a theory of massive gravity is of course much wider than pertains its use for late-time cosmology. Within string theory for example, open strings have spin-2 excitations whose lowest energy state is massive at tree level. The formulation of \cite{deRham:2010kj} in particular,  has found applications as varied as  e.g. its use as a framework for translational symmetry breaking and dissipation of momentum in holography \cite{Vegh:2013sk,Blake:2013owa}. 
Our interest is focussed on the inflationary context: here a consistent massive spin-2 field next to GR and a scalar inflaton field takes the form of a theory known as bigravity. This is an extension of dRGT theory that contains the same ghost-free structure. In bigravity each of the two metrics, $g$ and $f$, has its own Einstein-Hilbert term and they interact via the dRGT potential. The action reads
\bea\label{ggg}
S= \int {\rm d}^4 x \Bigg[ M_P^2\,\sqrt{-g}\,R[g] +\sqrt{-g}\, P_g(X,\varphi) +2 \sqrt{-g}\, m^2 M^2 V+  M_f^2\,\sqrt{-f}\,R[f] \Bigg], \,\,\,\,
\label{model}
\eea
where a few details are in order:

\begin{itemize}
\item The interaction potential $V$ is defined as
\begin{equation}
V=\sum_{n=0}^4\beta_n\,\mathcal{E}_n(\sqrt{g^{-1}f})\, ,
\end{equation}
where the $\beta_n$ are free parameters and the polynomials $\mathcal{E}_n(\mathbb{X})$ take the form
\bea
&&\mathcal{E}_0(\mathbb{X})=1, \quad \mathcal{E}_1(\mathbb{X})={\rm Tr}(\mathbb{X}) \equiv [\mathbb{X}], \quad \mathcal{E}_2(\mathbb{X})=\frac{1}{2}\left( [\mathbb{X}]^2 - [\mathbb{X}^2]  \right),\nonumber\\
&&\mathcal{E}_3(\mathbb{X})=\frac{1}{3!}\left( [\mathbb{X}]^3 -3 [\mathbb{X}^2][\mathbb{X}]+2 [\mathbb{X}^3]  \right), \nonumber\\
&&\mathcal{E}_4(\mathbb{X})=\frac{1}{4!}\left( [\mathbb{X}]^4 -6 [\mathbb{X}^2][\mathbb{X}]^2 +8 [\mathbb{X}^3][\mathbb{X}] +3 [\mathbb{X}^2]^2-6[\mathbb{X}^4]   \right)\, .
\label{poly}
\eea

\item  Due to the properties of the $\mathcal{E}_n$ polynomials, if it were not for the coupling to matter (here restricted to the metric ``g") the action would be symmetric under the exchange $g \leftrightarrow f\quad  M_g \equiv M_P \leftrightarrow M_f,\quad \beta_n \leftrightarrow \beta_{4-n} $.               
\item  The mass $M$ has already been symmetrized via
\be
M^2=\frac{M_P^2 M_f^2}{ M_P^2 +M_f^2}\,.
\ee
and we have introduced $\kappa\equiv\frac{M_f^2}{M_P^2}$.
\item As a consequence of the above points, it is not strictly true that $g$ is the massless spin-2 field and $f$ is the massive one. In fact, the mass eigenstate of the theory are in general time-dependent. We can on the other hand work in a low energy configuration \cite{DeFelice:2014nja} where this is approximately true.                        \\
\item The $P_g(X,\varphi)$ stands for a generic matter Lagrangian and {$X=-\frac{1}{2} g^{\mu \nu} \p_{\mu}\varphi \p_{\nu}\varphi$} . We will assume the potential within $P_g(X,\varphi)$ is driving the background dynamics of the inflaton field $\varphi$, minimally coupled to $g$ only. We refer the reader to \textit{Section \ref{matter}} for the more general case.                       
\end{itemize}

\noi The theory in Eq. (\ref{model}) has 8 propagating degrees of freedom: 5 (2T+2V+1S) from a healthy spin-2, 2 more from a massless spin-2 (2T) and a scalar. In what follows we will focus our attention on the $2 \times 2\,T$ degrees of freedom plus one scalar. Let us elaborate on the reasoning behind this choice. Non-linear massive (bi)gravity is endowed with a very efficient screening mechanism known as Vainshtein screening \cite{Vainshtein:1972sx,Babichev:2013usa}. In specific configurations the coupling to matter of the helicity-0 and helicity-1 modes is highly suppressed. This is, for example, what allows massive gravity to pass a host of cosmological tests in setups (e.g. solar system scales) where observations are in exquisite agreement with general relativity: the additional degrees of freedom are screened. We are working in a rather different, inflationary, background here; we will nevertheless assume that the non-linearities in the helicity-0 and helicity-1 part of the theory will generate sufficient screening so that we can focus the analysis on the remaining dofs. 

There exist in the literature several important studies \cite{Lagos:2014lca,Johnson:2015tfa,Akrami:2015qga,Cusin:2015pya,Brax:2017hxh} on the perturbation theory of all the degrees of freedom (including matter) populating bigravity during the various cosmological epochs. Requiring that an efficient screening is in place in an inflationary background is tantamount to requiring that non-linearities in the theory (these studies usually stop at quadratic order in perturbation theory) will downplay any effect due to the vector and helicity-0 modes.  
In support of this assumption is also the intuition that Vainshtein screening is typically more efficient in homogeneous and isotropic backgrounds. On the other hand, we stress here that studies of perturbations limited to quadratic order already indicate \cite{Cusin:2015pya} that all bigravity degrees of freedom are well-behaved during inflation.

The Vainshtein mechanism relies on the fact that whenever non-linearities are important the canonical normalization of e.g. the helicity-zero mode is redressed and so will be also the coupling with matter. The helicity-0 (usually indicated by $\pi$) coupling to matter is dictated by the structure in Eq.~(\ref{model}) and includes \footnote{Also terms such as $\frac{1}{\Lambda^3} \p_{\mu} \pi \p_{\nu} \pi\, T^{\mu \nu}$ are allowed, see \cite{deRham:2014zqa}. } the term $\pi\,T^{\mu}_{\mu}$. Whenever non-linearities in $\pi$ modify the normalization of the kinetic term, the corresponding coupling to matter will be 
\bea
 \pi_c\,T^{\mu}_{\mu}=  \frac{1}{Z(\pi_0, \beta_n)}\,\pi\,T^{\mu}_{\mu},
\label{mattercoupling}
 \eea
where $Z$ is a function of background quantities and the non-linearities are weighted by the $\beta_n$s. As soon as $\pi$ self-interactions are important, $Z\gg1$ and screening takes place. It is important to note at this stage that the same does not apply to the  tensor dofs introduced by making the metric $f$ dynamical. Even though $f$ does not directly couple to matter, it does so via $g$ and so the coupling does not undergo the same kind of suppression as the helicity-0/1 modes. On this basis we can go on and prioritize the observational effects of the tensor degrees of freedom contained in $f$. What must not be omitted in our analysis are the consequences that purely formal consistency criteria on the helicity-0 and vector sector have on the observables made up by the remaining dofs. The most important case in point is the helicity-0 unitarity bound, which we address next. The reader familiar with the matter and more interested in the calculation for the bispectrum signal may want to skip the content of the rest of \textit{Section \ref{model0}} and go directly to  \textit{Section \ref{calc}}. 

\subsection{The Higuchi bound and its observational consequences}
\label{higobs}
The unitarity bound on the helicity-0 mode corresponds to the condition on the coefficient of the kinetic term in the quadratic theory to be positive\footnote{See \cite{Kehagias:2017rpe} for a derivation of the Higuchi bound from the perspective of the AdS/CFT correspondence and in particular as a consequence of reflection positivity in radially quantized ${\rm CFT}_3$.}. This stems from requiring that the Hamiltonian is bounded from below so that the dynamical system is stable. Naturally, the latter also implies that the coefficients of the mass and spatial gradient terms, if present, need also be positive. It turns out that the helicity-zero Higuchi bound is the most restrictive on the parameter space of bigravity theory, so that it is justified to study it more in detail. It is well-known that unitary representations of the de Sitter group for spin-2 fields are massless, partially massless, and massive, with 
\bea
m^2=0\; {\rm (GR)},\qquad m^2=2 H^2\; {\rm (partially\, massless)} ,\qquad m^2 >2 H^2\; {\rm (massive)}\,.\quad
 \eea
It has been shown \cite{Fasiello:2013woa} that the existence of helicity-1 and helicity-0 interactions in the partially massless case implies the absence of the intriguing conjectured partially massless symmetry for both massive gravity and bigravity, at least as long as the kinetic term is of the standard form. In the massive representation, our stepping away from de Sitter and, most importantly, our dealing with (i) a fully non-linear theory of (ii) bigravity rather than massive gravity, relaxes \cite{Fasiello:2013woa} to some extent the bound on the now dressed mass $\tilde{m}$:  
\bea
\tilde{m}^2 \left( 1 + \frac{M_P^2 H_f^2}{M_f^2 H^2}\right) \geq 2H^2
\label{higuchi}
\eea
where we have defined $\tilde{m}^2$ as
\bea
\tilde{m}^2\equiv \frac{m^2}{2}\frac{H}{H_f}\frac{M^2}{M_P^2} \left(\beta_1 +2\beta_2 \frac{H}{H_f}+ \beta_3 \frac{H^2}{H_f^2} \right),
\label{mtilde}
\eea 
and the quantities $H$ and $H_f$ are the Hubble scales associated to the FLRW solution for the metrics $g$ and $f$ respectively\footnote{ Note that the absence of $\beta_0, \beta_4$ is due to the fact that these quantities are typically fixed by the tadpole cancellation requirement.}. These notions have immediate consequences also for the tensor sector and, as we shall see, for the observability of the $\langle \gamma\zeta \zeta \rangle $ bispectrum in the squeezed limit. In fact, we anticipate here (see Eqs.(\ref{uno})-(\ref{sol01}) in the text) that the mass in the (suitably diagonalized) tensor equations of motion in the low $k$ limit is the very same LHS of Eq. (\ref{higuchi})  \cite{Comelli:2012db}. This implies that the helicity-0 mode unitarity bound forces an effective mass on the tensor sector, and in particular one of at least order Hubble: the corresponding tensor wavefunction will then inevitably have a component that decays outside the horizon, to the detriment of the observed signal. As for possible tachyonic and gradient instabilities, we will work in a regime where any such possibility is above the reach of the effective theory: the ``low energy" regime of \cite{DeFelice:2014nja}.

\subsection{Strong coupling scales}
\label{scs}
For the sake of providing a more complete picture, we find it necessary to briefly expound on the strong coupling scale (scs) of the bigravity theory at hand. Naturally, it is the presence of the $g-f$ interaction term that dictates the lowest scales in the theory. In particular, depending on the hierarchy between the two masses $M_g$ and $M_f$, the lowest ``naive" strong coupling scale (in the helicity-0 sector) can be as low as: 
\bea
\Lambda^{g,f}_3 = \left(m^2 M_{g,f}\right)^{1/3}\; .
\eea
A few comments are in order.\\
Taking at face value the above quantities may lead to some worries about a scs perilously close to the energy we are working at. Indeed, although the Higuchi bound forces the dressed mass to be larger than the Hubble scale $H$, one typically satisfies this condition by requiring $H_f/M_f\gg H/M_g$, and allowing $m^2\lesssim H^2$. This will be also the case for our parameter space of choice as keeping the ratio $M_f^2/M_g^2\cdot H^2/H^2_f$ small will allow approximately constant mass eigenstates in the tensor sector  \cite{DeFelice:2014nja}. However, the scs is naive in the sense that the background via the Vainshtein mechanism changes the canonical scale
and so the scale at which the fluctuations are strongly coupled. Instead of the naive scale $\Lambda$, one should rather think of a larger quantity as the strong coupling scale, $\Lambda^* = \sqrt{Z} \Lambda$, where $Z$ is the same quantity discussed below equation (\ref{mattercoupling}). The efficient screening we require goes then in the same direction of a larger strong coupling scale.

\subsection{Coupling to matter}
\label{matter}
In writing Eq. (\ref{model}) we have assumed the existence of a unique matter sector that  is minimally coupled to the metric $g$ only; it is in fact this coupling (together with the hierarchy between $M_g$ and $M_f$) that breaks the otherwise symmetric role played by the two metrics in our setup. One might ask: is this the only possibility? Would any other coupling lead to a ghostly (classical or quantum) degree of freedom? How about coupling different matter sectors to different metrics? Naturally, coupling   matter just to the $f$ sector is just as acceptable but the $g\leftrightarrow f$ exchange would be mirrored by a swapping of the observables we are going to calculate, and deliver no detectable advantage. Coupling different matter sectors to different metric fields is also not immediately helpful because communication between different matter sectors will be mediated and therefore suppressed. On the other hand, what might prove to be more intriguing for us is the coupling via the effective composite metric \cite{deRham:2014naa}:
\bea
g^{\rm eff}_{\mu\nu}=\alpha^2 g_{\mu \nu} + 2\alpha\, \beta\, g_{\mu\alpha} \left(\sqrt{g^{-1} f}   \right)^{\alpha}_{\nu}+\beta^2 f_{\mu\nu}\;,
\label{compositem}
\eea 
where $\alpha$ and $\beta$ are two arbitrary real dimensionless parameters. In this case the matter Lagrangian in Eq. (\ref{model}) would be replaced,  in a simple example, by:
\bea
\mathcal{L}_{m}= \sqrt{-g_{\rm eff}} \Big[g^{\mu\nu}_{\rm eff}\,\p_{\mu}\varphi\p_{\nu}\varphi+V(\varphi)  \Big].
\label{compositeL}
\eea
This specific form for the coupling preserves the ghost-free nature of the theory both at the classical and quantum level \cite{deRham:2014naa}. From the EFT perspective any ghost that is above the cutoff scale of the theory is completely harmless and it is in this sense that one can freely use the coupling via composite metric in Eq.~(\ref{compositem}).

\section{Non-standard tensor-scalar-scalar consistency relations}
\label{calc}

Equipped with sufficient information on our bigravity model, we now go on to tackle the observational signal generated by the tensor-scalar-scalar three-point function $\langle\gamma_g\zeta\zeta\rangle$, where $\gamma_g$ is the $g$-metric tensor fluctuation and $\zeta$ is a repository of scalar fluctuations. As mentioned in \textit{Section \ref{intro}} and detailed in \textit{Appendix \ref{crs}}, non standard CRs are possible whenever additional field content has non-linear transformation under a gauge diff. The extra field content of choice here includes the $f$-metric new tensor dof. In order to most clearly show how consistency relations are modified in our setup we provide in this manuscript two different routes to non standard CRs. We present the first one below, in \textit{Section \ref{route2}}, and leave the second one to \textit{Appendix \ref{route1}} . The latter relies on an approximation that 
does not hold outside the horizon and cannot therefore be trusted all the way to late-time observables. Nevertheless, we find the presentation in \textit{Appendix \ref{route1}} very instructive and choose to include it in the manuscript. In \textit{Section \ref{route2}} we will solve the full\footnote{This is in contradistinction with the procedure outlined in \textit{Appendix \ref{route1} } where an approximation of the type $\delta \mathcal{L}^{(2)} \ll \mathcal{L}^{(2)}_0$ on parts of the quadratic Lagrangian is necessary.} coupled system of equations for the traceless transverse part of $h_f,h_g$.   We will then calculate the tree level $\langle\gamma_g \zeta \zeta\rangle$ diagram and show that its late time limit is different than that of the standard single-field model as it contains a term that depends on the mass $m$. The late-time power spectra $P_{\gamma_g}$\footnote{We do not include $P_{\gamma_f}$ here because $\gamma_f$ does not couple directly with matter. However, the conclusion would not change as at tree level and in the late time limit $P_{\gamma_f}$ too is unaware of the mass $m^2$.}  and $P_{\zeta}$ cannot capture this mass dependence at tree level, signaling that CRs are indeed modified in our setup.

\subsection{Background}
Following for example \cite{Comelli:2012db,DeFelice:2014nja,Fasiello:2013woa}, it is straightforward to derive background equations and study the quadratic perturbation theory around FLRW solutions for the metric $f$ and $g$ . We denote
\bea
\bar{g}_{\mu\nu}\, {\rm d}x^{\mu}{\rm d}x^{\nu}= -N^2 {\rm d}t^2 + a^2 \delta_{ij}\,{\rm d}x^i {\rm d}x^j, \quad \bar{f}_{\mu\nu}\, {\rm d}x^{\mu}{\rm d}x^{\nu}= -\tilde{N}^2 {\rm d}t^2 + b^2 \delta_{ij}\,{\rm d}x^i {\rm d}x^j\; .
\eea
Here $N$ and $\tilde N$ are background lapse functions and $a$ and $b$ are the scale factors for the corresponding metrics $g$ and $f$.
The Friedmann equations read:
\bea
&&3 M_P^2 H^2= \rho(\varphi) + 3 m^2 M^2 \sum_{n=0}^3 \frac{\beta_n}{(3-n)! n!}\left(\frac{b}{a} \right)^n \;,
\label{frhg}
\eea
\bea
&&3 M_f^2 H_f^2=  3 m^2 M^2 \sum_{n=0}^3 \frac{\beta_{n+1}}{(3-n)! n!}\left(\frac{b}{a} \right)^{n-3} \;,
\label{frhf}
\eea
where $\rho(\varphi)$ in Eq.~(\ref{frhg}) and in particular the potential $V(\varphi)$ in it, is the leading term, driving inflation, on the RHS. {We notice in passing that we will be working in the so-called healthy branch (see e.g. \cite{DeFelice:2014nja}) of bigravity solutions in order to avoid strong coupling issues. The healthy branch is defined by the condition $H/H_f= b/a\equiv \xi$.\footnote{To conform with standard notation we have defined the ratio of the two scale factors as $\xi\equiv b/a$. This is also the symbol generally used for gauge transformations. Which one is meant should be clear from the context.}} Upon using 
\bea
g_{\mu\nu}=\bar{g}_{\mu\nu}+a^2\, h_{g\,\mu\nu}\; , \quad f_{\mu\nu}=\bar{f}_{\mu\nu}+b^2\, h_{f\,\mu\nu}\; ,
\label{perth}
\eea
one obtains from the quadratic action the equations of motion for $h_f, h_g$.
 
\subsection{Linear perturbations}
\label{route2} 
Let us start with the equations of motion for $\gamma_g^{TT}$ and $\gamma_f^{TT}$, the traceless transverse components of, respectively, $h_{g}$ and $h_{f}$ \footnote{As anticipated, henceforth we shall focus on the tensor sectors perturbations. For a systematic study of quadratic perturbations in bigravity, we refer the reader to, for example, the work in \cite{Comelli:2012db}.}. Tensor perturbations of a given k-mode are separable into two independent helicity modes with the same mode-function: $\gamma_{g},\,\,\gamma_{f}$. Combining Eq.~(\ref{model}) and (\ref{perth}) one obtains\cite{Comelli:2012db}:

\begin{align} \label{uno}
& \gamma_{g}'' -\frac{2}{\tau} \, \gamma_{g}^{'}+k^2 \gamma_{g}+\frac{\mathcal{M}^2}{H^2 \tau^2}\,\left(\frac{r_{fg}^2}{1+r_{fg}^2}\right)\,\left(\gamma_{g}- \gamma_{f}\right)=0,\\
& \gamma_{f}'' -\frac{2}{\tau} \, \gamma_{f}^{'}+k^2 \gamma_{f}+\frac{\mathcal{M}^2}{H^2 \tau^2}\, \frac{1}{\left(1+r_{fg}^2\right)}\,\left(\gamma_{f}- \gamma_{g}\right)=0, \label{due}
\end{align}
where $\mathcal{M}^2= \tilde{m}^2 \left[1+ 1/r_{fg}^2\right]$ and we introduced the parameter
\begin{equation}
r_{fg}\equiv \frac{M_f}{M_P}\frac{H}{H_f}\,.
\end{equation}
In writing Eqs.~(\ref{uno})-(\ref{due}) we have also assumed a slowly varying $\xi= b/a$ . It turns out to be convenient at this stage to change basis, from $\gamma_{g,f}$ to $\gamma_{+,-}$ , where

\begin{equation}
\gamma_{\pm}^{}=\gamma_{g}^{}\pm r_{fg}^2\, \gamma_{f}^{}\,,
\end{equation}
so that the equations of motion now read

\begin{align}\label{sol0}
& \gamma_{+}^{''}-\frac{2}{\tau}\,\gamma_{+}^{'}+k^2 \gamma_{+}=0,\\\label{sol01}
&\gamma_{-}^{''}-\frac{2}{\tau}\,\gamma_{-}^{'}+\left(k^2+\frac{\mathcal{M}^{2}}{H^2\, \tau^2} \right)\gamma_{-}-\frac{\mathcal{M}^{2}}{H^2\, \tau^2}\,\left(\frac{1-r_{fg}^2}{1+r_{fg}^2}\right)\,\gamma_{+}^{}=0\,.
\end{align}

\noi The solutions, once written back in the $\gamma_g, \gamma_f$ basis are
\begin{eqnarray} \label{sol1} 
\gamma_{g}&=&\left(\frac{1+r_{fg}}{1+r_{fg}^{2}} \right)\frac{H}{M_{P}}\frac{\sqrt{2} \left(i-k\tau\right)}{k^{3/2}}e^{-ik\tau} \\&&+  \left(\frac{-1+r_{fg}}{1+r_{fg}^2} \right)r_{fg}\,\frac{H}{M_{P}} e^{i\nu\pi} (1+i)\sqrt{\frac{\pi}{2}}\left(-\tau\right)^{3/2}H^{(1)}_{\nu}(-k\tau) \nonumber   \,,\,\,\\\label{sol2}
\gamma_{f}&=&\left(\frac{1+r_{fg}}{1+r_{fg}^2} \right)\frac{H}{M_{P}}\frac{\sqrt{2} \left(i-k\tau\right)}{k^{3/2}}e^{-ik\tau} \\&&+  \left(\frac{1-r_{fg}}{1+r_{fg}^2} \right)\frac{1}{r_{fg}}\frac{H}{M_{P}} e^{i\nu\pi} (1+i)\sqrt{\frac{\pi}{2}}\left(-\tau\right)^{3/2}H^{(1)}_{\nu}(-k\tau)  \nonumber  \,,\,\, 
\end{eqnarray}

\noi where we have defined  $\nu\equiv \sqrt{9/4-\mathcal{M}^2/H^2}$.\\    
\noi Note that the solution for the mode-functions in Eqs.~(\ref{sol1})-(\ref{sol2}) were found by solving Eqs.(\ref{sol0})-(\ref{sol01}) first and requiring that the mode functions asymptote to Bunch-Davies vacua deep inside the horizon:\bea
\gamma_{g}(k,\tau)\rightarrow_{_{|k\tau|\gg1}} \frac{2}{a\,M_{P}}\,\frac{e^{-i\,k\,\tau}}{\sqrt{2k}}\,,\qquad \gamma_{f}(k,\tau)\rightarrow_{_{|k\tau|\gg1}} \frac{2}{b\,M_{f}}\,\frac{e^{-i\,k\,\tau}}{\sqrt{2k}}\,. 
\eea
 It is worth mentioning the special case $r_{fg}^2=1$. In this particular configuration, despite the presence of the mass term in the equations of motion, the two solutions for $\gamma_g, \gamma_f$ are independent and identical to two copies of the usual massless solution for $\gamma_g$ in standard single-field inflation. One concludes that in this case, at the level of the quadratic action for the traceless transverse tensor components, the theory behaves as two copies of GR and one has to go to cubic order to start probing the massive gravity interactions.\\
\subsection{Tensor-scalar-scalar bispectrum}

Equipped with the solution to the quadratic theory, we can now consider the effect of the $m^2$ driven $g-f$ coupling on standard observables such as the tensor-scalar-scalar bispectrum.  We shall be considering the contribution due to the $ \gamma_g^{ij} \p_i \zeta \p_j \zeta$ interaction. Note that we define the gauge-invariant comoving curvature perturbation $\zeta$ in the standard way, $\zeta=\psi_g+{\cal H}\delta\rho/\bar{\rho}'$, where
$\psi_g$ is the gravitational potential identified with the diagonal part of the spatial components of the metric $g$ (see also \cite{Lagos:2014lca,Cusin:2015pya}). The important question is if this quantity is conserved on scales larger than the Hubble radius. Since it is only the metric $g$ that interacts with matter, in light of the conservation equation and of our assumption that the helicity-0 and helicity-1 modes are screened, one immediately finds that $\zeta'=\psi'_g=0$ outside the horizon, and we conclude that one can adopt for $\zeta$ the usual expression. This is consistent with e.g. the analysis in \cite{Cusin:2015pya}.
Since the $ \gamma_g^{ij} \p_i \zeta \p_j \zeta$ interaction is already present in single-field inflation, we can already anticipate that the imprint of the massive gravity potential is all stored in the wavefunctions and, in particular, their $\mathcal{M}$-dependent parts. 
\noindent The formal expression for the bispectrum in the in-in formalism is
\begin{equation}
\Big\langle \gamma_{g}^{\lambda}(t,\vec{k}_{1})\zeta(t,\vec{k}_{2})\zeta(t,\vec{k}_{3})\Big\rangle =-i\int_{t_{0}}^{t} {\rm d}t' \Big\langle \left[\gamma_{g}^{\lambda}(t,\vec{k}_{1})\zeta(t,\vec{k}_{2})\zeta(t,\vec{k}_{3}),H_{\text{int}}(t')\right]\Big \rangle
\end{equation}
where 
\bea
H_{\text{int}}(t')=-\mathcal{L}_{\text{int}}(t')=-M_{P}^{2}\epsilon\int {\rm d}^{3}x\, a(t')\,\gamma_{g,ij}^{}(\vec{x},t')\partial_{i}\zeta(\vec{x},t')\partial_{j}\zeta(\vec{x},t')\, .
\label{interaction}
\eea

\noi It is convenient to split the result of the bispectrum calculation in two pieces. One, which we indicate as $\mathcal{B}_{\text{\sout{m}}}$, will be due to the ``massless"\footnote{The massless component of $\gamma_g$ is the first term in Eq.~(\ref{sol1}), which is unaware of $\nu=\nu(\mathcal{M})$ and is therefore termed  massless.  } component of $\gamma_g$ in the integral of Eq.~(\ref{interaction}); the other one, corresponding to the remaining component  of the wavefunction, will be indicated by $\mathcal{B}_{\text{{m}}}$. One obtains:   
\begin{eqnarray}
\Big\langle \gamma_{g}^{\lambda}(t,\vec{k}_{1})\zeta(t,\vec{k}_{2})\zeta(t,\vec{k}_{3})\Big\rangle =(2\pi)^3\delta^{(3)}(\vec{k}_{1}+\vec{k}_{2}+\vec{k}_{3})e_{ij}^{\lambda}(\hat{k}_{1})\left(-\hat{k}_{2i}\hat{k}_{3j}-\hat{k}_{3i}\hat{k}_{2j}\right)\left[\mathcal{B}_{\text{\sout{m}}}^{}+\mathcal{B}_{\text{m}}^{}\right]\,.\nonumber\\
\end{eqnarray}
It is straightforward (and identical to the single-field case) to derive the quantity in $\mathcal{B}_{\text{\sout{m}}}$. The integral necessary to obtain an explicit expression for $\mathcal{B}_{\text{m}}$  can be performed analytically by fixing each time the value of the mass $\mathcal{M}$. Let us fix it to $\mathcal{M}=\sqrt{2}\,H$, a value that saturates the Higuchi bound. The explicit expression for the bispectrum now reads: 
\begin{eqnarray}\label{resb1}
&&\mathcal{B}_{\text{\sout{m}}}^{} (k_1,k_2,k_3)= \frac{H^4}{4\,\epsilon\,M_{P}^{4}}\, \left(\frac{1+r_{fg}}{1+r_{fg}^2} \right)^2\mathcal{I}_{\text{\sout{m}}}(k_1,k_2,k_3) \,,\\\label{resb2}
&&\mathcal{B}_{\text{m}}^{} (k_1,k_2,k_3)= \frac{H^4}{4\,\epsilon\, M_{P}^{4}}\frac{1-r_{fg}^2}{\left(1+r_{fg}^2 \right)^2}\,r_{fg}\,\mathcal{I}_{\text{m}}(k_1,k_2,k_3)  \,,
\end{eqnarray}
where
\begin{eqnarray}
&& \mathcal{I}_{\text{\sout{m}}}(k_1,k_2,k_3)\equiv \frac{\sum_{i}k_{i}^{3}+2\sum_{i\not= j}k_{i}k_{j}^{2}+2 \Pi_{i}k_{i}}{k_{1}^{3}k_{2}^{2}k_{3}^{2} k_{t}^{2}}\,,\\
&& \mathcal{I}_{\text{m}}(k_1,k_2,k_3)\equiv \frac{1}{\Pi_{i}k_{i}^{2}}\left(\frac{k_1 (k_2+k_3)+3k_2 k_3+k_2^2+k_3^2}{k_{t}^{2}}-\gamma-\ln\left[-k_{t}\tau_{0}\right]\right)\,,\nonumber\\
\end{eqnarray}
with $k_{t}\equiv k_1+k_2+k_3$ and where $\gamma$ is the Euler's constant. The logarithm in the last expression is handled in the standard fashion \cite{Riotto:2008mv,Burgess:2009bs,Seery:2010kh}. The presence of a log term in the bispectrum signals a non-trivial interaction between the hard $\zeta$ modes and the soft graviton outside the Hubble radius and accounts for the
fact that  the massive graviton mode, despite its decay on large scales, sources the massless graviton perturbation from the moment it leaves the Hubble radius till the end of inflation. Such an effect is not  found in the standard single-field scenario.\\ We can now take the squeezed limit ($k_{1}\ll k_{2}\simeq k_{3}$) of the expressions in (\ref{resb1})-(\ref{resb2}), to get
\begin{align}\label{resb11}
\mathcal{B}_{\text{\sout{m}}}^{} (k_1,k_2,k_3) &\overset{k_{1}\ll k_{2}\simeq k_{3}}{\xrightarrow{\hspace*{1cm}}} \frac 34 P_{\gamma_g}^{\lambda}(k_1) \,P_\zeta(k_3) ,\\\label{resb21}
\mathcal{B}_{\text{m}}^{} (k_1,k_2,k_3) &\overset{k_{1}\ll k_{2}\simeq k_{3}}{\xrightarrow{\hspace*{1cm}}} \frac 18 P_{\gamma_g}^{\lambda}(k_1) \,P_\zeta(k_3)\,\frac{k_1}{k_3}\,\left(\frac{1-r_{fg}}{1+r_{fg}}\right) \,r_{fg} \left(5-4\gamma-4\ln[-2k_3\tau_0]\right),
\end{align}
where we have defined the tensor and scalar power spectra at late times\footnote{Note that the second term in the mode-function for $\gamma_g$, Eq. \eqref{sol1}, generates terms scaling as positive powers of $k \tau$ for $k\tau\rightarrow0$, therefore they do not contribute to the power spectrum at late times.}
\begin{align}\label{ts}
P_{\gamma_g}^{\lambda}(k_1)&= \lim_{\tau\rightarrow 0}\left[\gamma_{g}(k_1,\tau)\gamma_{g}^{*}(k_1,\tau)\right]\simeq \left(\frac{1+r_{fg}}{1+r_{fg}^2} \right)^2 \frac{2\,H^2}{M_{P}^{2}}\frac{1}{k_{1}^{3}}\,,\\\label{cs}
 P_{\zeta}(k_{3})&=\lim_{\tau\rightarrow 0}\left[\zeta(k_3,\tau)\zeta^{*}(k_3,\tau)\right]\simeq\frac{H^2}{4\,\epsilon\,M_{P}^{2}}\frac{1}{k_{3}^{3}}\,
\end{align}
A tensor-scalar-scalar bispectrum  that satisfies the consistency condition would, in the squeezed limit, behave according to
\be
\langle \gamma_{g}^{\lambda}(\vec{k}_{1})\zeta(\vec{k}_{2})\zeta(\vec{k}_{3})\overset{k_{1}\ll k_{2}\simeq k_{3}}{\xrightarrow{\hspace*{1cm}}} -(2\pi)^3\delta^{(3)}(\vec{k}_{1}+\vec{k}_{2}+\vec{k}_{3})\frac{1}{2}e_{ij}^{\lambda}(\hat{k}_{1})\hat{k}_{3i}\hat{k}_{3j}P_{\gamma_{g}}^{\lambda}(k_{1})P_{\zeta}(k_3)\frac{\partial\ln P_{\zeta}(k_3)}{\partial\ln k_3}\,.
\label{usual}
\ee
As expected, the contribution from $\mathcal{B}_{\text{\sout{m}}}^{}$ in (\ref{resb11}) does indeed obey such rule. The squeezed limit contribution from $\mathcal{B}_{\text{{m}}}$ in (\ref{resb21}), however, does not: its expression depends on $\mathcal{M}$ whilst both the tree level scalar and tensor power spectrum do not. The $\mathcal{M}$ dependence in Eq. (\ref{resb21}) is implicit in that the exponent of the very last term, $k_1/k_3$, can be thought of as proportional to $3/2-\sqrt{9/4-\mathcal{M}^2/H^2}$. {We stress that this is specific to the case $\mathcal{M}=\sqrt{2}\,H$ and in contradistinction to the case where a similar (except for an angular dependence) suppression arises from a different diffeomorphism altogether in standard single-field inflation where CRs are conserved \cite{Creminelli:2012ed}.}  Note also that we have not included the contribution from $\gamma_f$ to the late time tensor power spectrum. This is because this contribution is suppressed in light of $\gamma_f$ not coupling directly to matter. Nevertheless, our conclusions would not change as the late-time limit of $\gamma_f$ is unaware of $\mathcal{M}$ and therefore cannot cancel off the $\mathcal{B}_{\text{{m}}}$ bispectrum contribution in the squeezed limit. As we shall see more at length in \textit{Section \ref{obs}}, the small ratio $k_1/k_3$ in Eq. (\ref{resb21}) and the proportionality to $r_{fg}$ play a key role in determining the detectability of the observational signal. As we have remarked above, a smaller $\mathcal{M}$ would lead to a smaller exponent for $k_1/k_3$ and thus a larger signal. We shall now elaborate on different strategies, within the same model, to enhance the signal.
\subsection{All of the freedom in the unitarity bound}
The Higuchi bound we inherit from the helicity-0 mode is strictly the one in Eq. (\ref{higuchi}) only in the low $k$ limit, at or outside the horizon, in inflationary terms. From Ref. \cite{Comelli:2012db} we know that the scalar sector of the perturbations develops an instability (see Fig.~\ref{fig5}) unless

\bea
\frac{4/3\, k^6 + 6 k^4 \mathcal{H}^2}{9a^2 \mathcal{H}^2 \mathcal{M}^2+k^4 -18 \mathcal{H}^4 }-\frac{12 k^8 \mathcal{H}^2 }{(9a^2 \mathcal{H}^2 \mathcal{M}^2+k^4 -18 \mathcal{H}^4 )^2}+ a^2 \mathcal{M}^2 -\frac{k^2}{3} -2\mathcal{H}^2 \geq 0\;, \; \qquad
\label{ghiguchi}
\eea
where $\mathcal{H}=a H$. The above implies that for, $k^2> 8\, a^2 H^2$, there is effectively no bound on $\mathcal{M}$. In other words, up to what corresponds approximately to the horizon for the long mode, $\mathcal{M}$ can be much smaller than $H^2$ so that the corresponding exponent of the ratio $k_1/k_3$   would be much smaller and the corresponding signal in Eq. (\ref{resb21}) enhanced. 
For this to be true we would need a time-dependent $\mathcal{M}$, something not allowed in de Sitter but possible in quasi-dS. The crucial point however is that, upon studying the $k$ dependence of Eq. (\ref{ghiguchi}), one can see immediately that the inequality abruptly requires $\mathcal{M}>2H^2$ before we cross the $k^2=a^2 H^2$ threshold, with $k$ being the long mode. This implies that a small and time-dependent $\mathcal{M}$ would be in place only when short modes are still well inside their horizon and the long mode is about to exit. It is well-known that in this configuration the contribution from the integral in Eq. (\ref{interaction}) is in fact suppressed. We conclude then that a suitably adjusted time-dependent $\mathcal{M}$ can do little to improve the signal. 
 
\begin{figure*}[h]
\centering
\resizebox{0.47\textwidth}{!}{\includegraphics{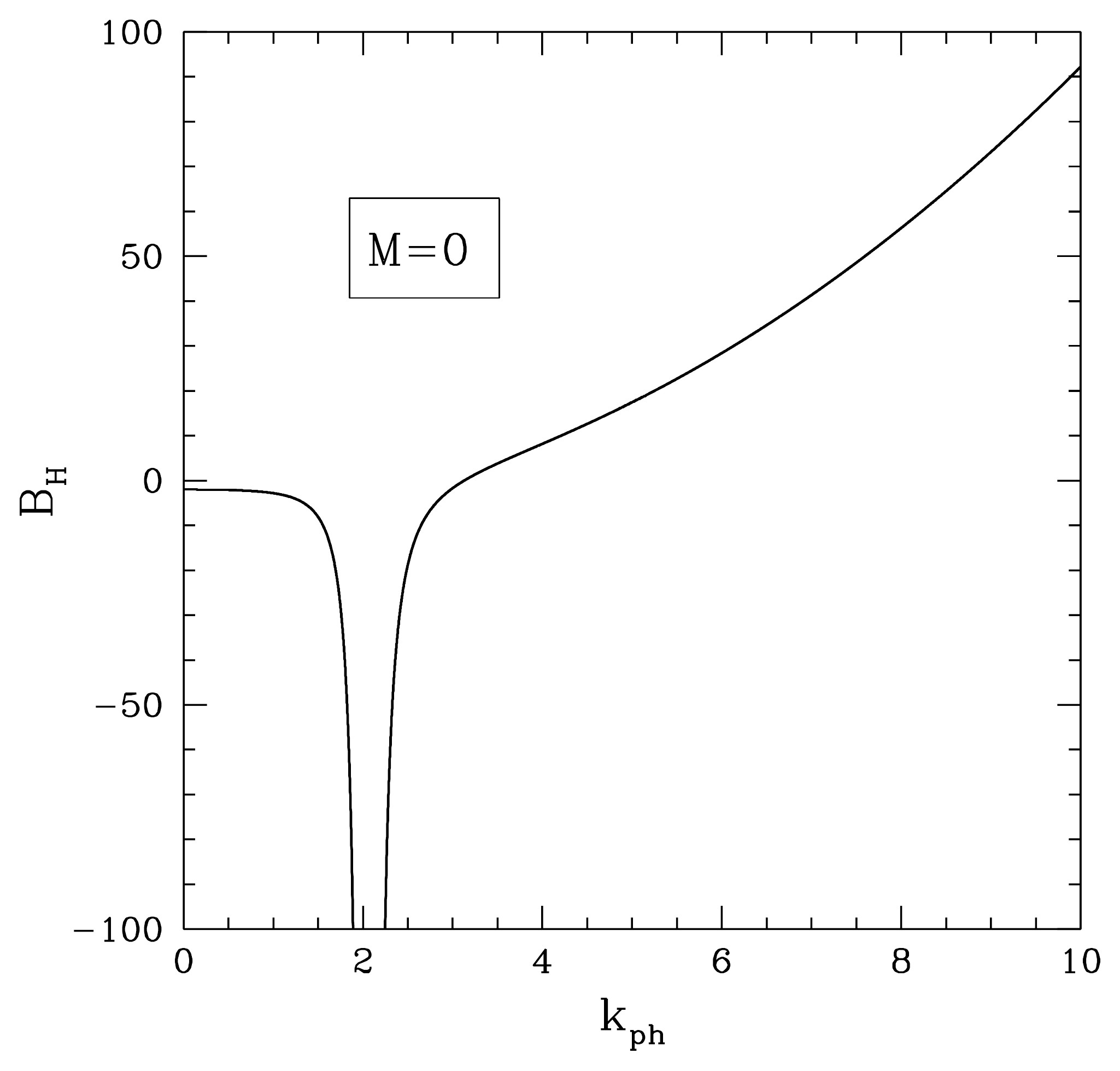}}
\resizebox{0.46\textwidth}{!}{\includegraphics{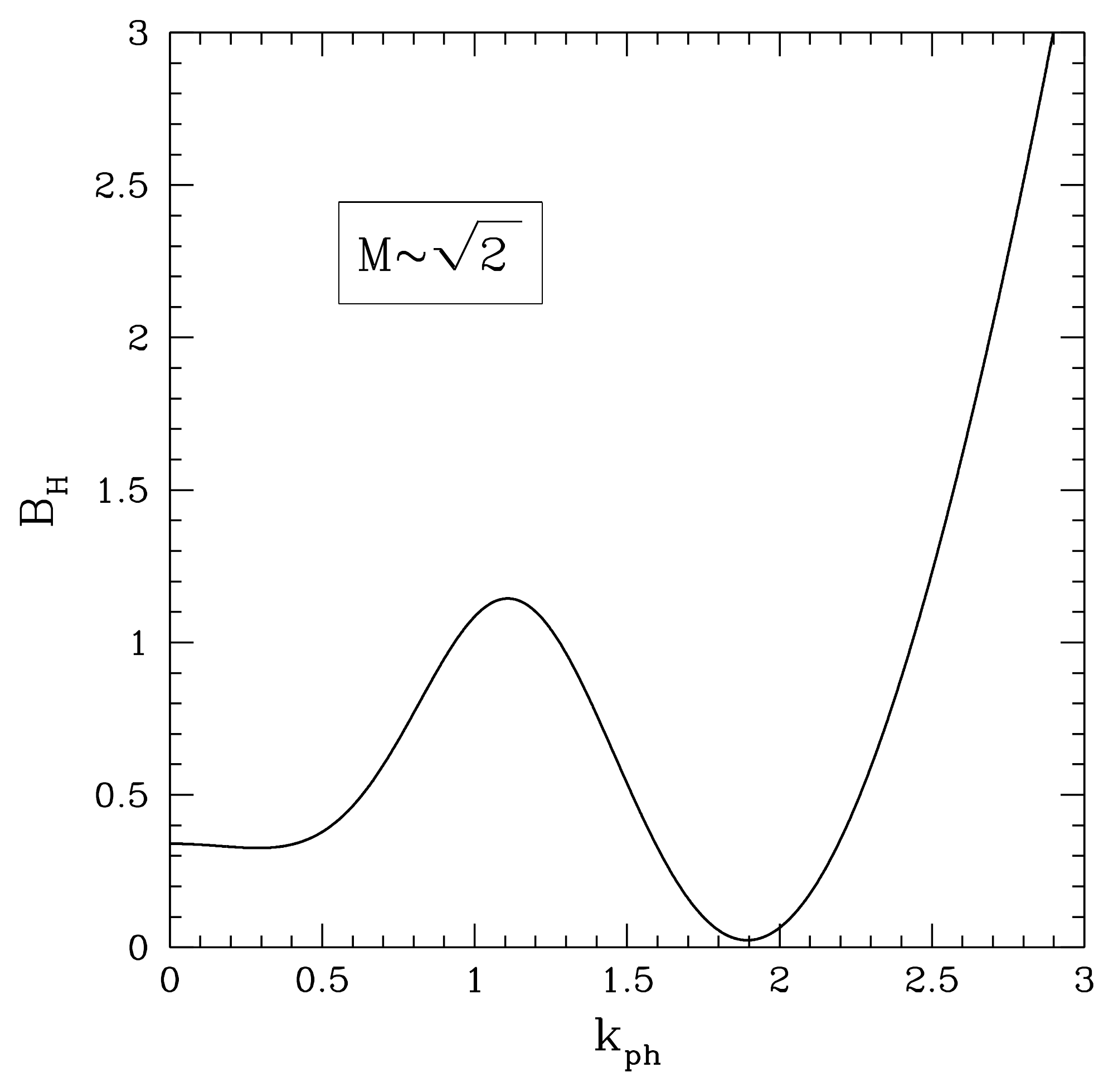}}
\caption{Left: the LHS of Eq. (\ref{ghiguchi}) as a function of the physical $k$ for $\mathcal{M}=0$. One can notice that a $k_{\rm ph} \geq \sqrt{8} H$ is enough for a positive $B_{H}$, regardless of the value of $\mathcal{M}$. Right: the LHS of Eq. (\ref{ghiguchi}) as a function of the physical $k$ for $\mathcal{M}\sim \sqrt{2}$; for small $k$ the value of $\mathcal{M}$ needs to surpass this threshold in order for the Higuchi bound to be satisfied.} 
\label{fig5}
\end{figure*}

\subsection{New cubic interactions}
As stated, the $k_1/k_3$ suppression in Eq. (\ref{resb21}) is due to the non-zero $\mathcal{M}$-dependent term in the tensor wavefunction. Of course, simply considering the wavefunction massless component leads to the standard CR. The next logical step is to consider the massless component of the wavefunction and switch on new cubic interactions originating from the massive gravity potential $\mathcal{E}(\sqrt{g^{-1}f})$. In light of the fact that these interactions diagram(s) will be inserted within the $\langle \gamma\zeta\zeta \rangle$ three-point function, it is clear that we will be dealing with one-loop diagrams. Loops would typically lead to a $(H/M_P)^2$ suppression but in bigravity we have a little more freedom in the form of the  $M_P/M_f$ hierarchy. \\
\indent To ensure non-standard CRs, we also need to verify that the resulting squeezed bispectrum cannot be expressed as in Eq. (\ref{usual}) by considering the corresponding possible one-loop contributions in $P_{\gamma}$ or $P_{\zeta}$.  The diagrams below should be considered as indicative of the typical behaviour in our setup. In view of the general result, it is not necessary to calculate the exact numerical value of the interactions and it shall suffice to give an estimate. We will also temporarily relax the assumption that matter couples only to the metric $g$, we do so relying on couplings such as those in Eq. (\ref{compositem}). We resort here to direct coupling of $f$ to matter because it allows us to construct diagrams involving the new spin-2 degrees of freedom, have $\gamma_g$ and two $\zeta$s as external fields, and stay at one loop order. Two-loops would be necessary without direct coupling between $f$ and matter. One possible diagram is shown in Fig.~\ref{fig5}.

\begin{figure*}[h]
\centering
\includegraphics[scale=0.55]{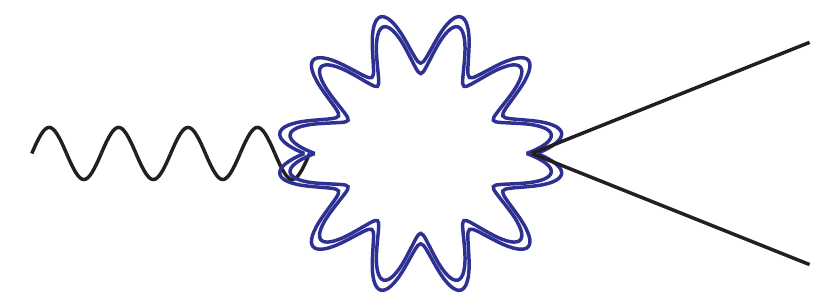}
\caption{Diagrammatic representation of a typical 1-loop interaction: -- the left vertex should be thought of as originating from a ``new" $m^2M^2$-type interaction; -- the right vertex stems from standard-type direct minimal coupling of $f$ with matter. The black (blue) wiggly line represents $\gamma_g$($\gamma_f$), the solid ones stand for $\zeta$.} 
\label{fig5}
\end{figure*}

\noi The first vertex from the left should be though of as originating from the new $m^2 M^2$-proportional interactions whilst the second vertex is ``standard" whenever matter is coupled directly to $f$ as well. A quick estimate of the amplitude of this contribution returns 
\bea
\mathcal{B}_{\text{\sout{m}}}^{} \sim \frac{H^4}{\epsilon_f M_P^4 }\cdot \frac{H^2}{M_P^2}
\eea
where we have imposed a $H\sim H_f, \quad M_f\ll M_P$ hierarchy. This result is to be compared with Eq. (\ref{resb21}). As expected, employing the massless components of the wavefunctions guarantees the absence of the $k_1/k_3$ factor. However, the price to pay is a Planck suppression squared. One can check that, regardless of the hierarchy, a suppression at least as strong as the typical loop scaling remains. Notice also that the outcome does not qualitatively change for the special configuration $r_{fg} =1$ discussed in \textit{Section \ref{route2}} .

\section{Observational consequences}
\label{obs}

\noindent If a coupling exists between a long-wavelength tensor and two short-wavelength scalar fluctuations, different scalar Fourier modes will appear to be correlated with one another. These \textsl{off-diagonal} correlations can be employed to estimate the amplitude of the tensor fluctuation. Note that such an observable   is only sensitive to sub-horizon tensor modes, as it is not possible to resolve scalar modes that differ by an amount smaller than $H_{0}^{-1}$. An optimal unbiased estimator for the tensor modes amplitude built from off-diagonal correlations was proposed in \cite{Jeong:2012df,Dai:2013ikl}, it reads:
\begin{equation}\label{set}
\hat{A}_{\gamma}=\sigma_{\gamma}^{2}\sum_{\vec{K},\lambda}\frac{\left(P^{\lambda}_{f}(K)\right)^{2}}{2\left(P^{\lambda}_{n}(K)\right)^{2}}\left(\frac{|\hat{\gamma}^{\lambda}(\vec{K})|^{2}}{V}-P^{\lambda}_{n}(K)\right)\, ,
\end{equation}
where $V\equiv(2\pi/k_{\text{min}})^{3}$ is the survey volume, $P_{f}^{\lambda}\equiv P_{\gamma}^{\lambda}(K)/A_{\gamma}$ is a fiducial power spectrum for the tensor modes, $\hat{\gamma}^{\lambda}(\vec{K})$ is the optimal estimator for a Fourier mode amplitude defined as
\begin{eqnarray}
\hat{\gamma}^{\lambda}(\vec{K})\equiv P^{\lambda}_{n}(K)\sum_{\vec{k}}\frac{B^{\lambda}(\vec{K},\vec{k},\vec{K}-\vec{k})/P_{\gamma}(K)}{2VP_{\rm tot}(k)P_{\rm tot}(|\vec{K}-\vec{k}|)}\delta(\vec{k})\delta(\vec{K}-\vec{k})\, .
\label{gamma3}
\end{eqnarray}
In Eq. (\ref{gamma3}) $P_{n}^{\lambda}$ stands for the  variance
\begin{eqnarray}\label{riprendi}
P_{n}^{\lambda}(K)\equiv \left[\sum_{\vec{k}}\frac{|B^{\lambda}(\vec{K},\vec{k},\vec{K}-\vec{k})/P_{\gamma}(K)|^{2}}{2VP_{\rm tot}(k)P_{\rm tot}(|\vec{K}-\vec{k}|)}\right]^{-1}\,.
\end{eqnarray}
The quantity $P_{\rm tot}$ is the total scalar power spectrum (signal$+$noise) and $B^{\lambda}$ is the effective (i.e. after subtraction of the consistency-conditions-preserving part) tensor-scalar-scalar bispectrum. The variance associated with (\ref{set}) is given by
\begin{equation}
\sigma^{-2}_{\gamma}=\frac{1}{2}\sum_{\vec{K},\lambda}\left[ K^{3} P^{\lambda}_{n}(K)\right]^{-2} \,\, ,
\end{equation}
and can be used to quantify, for a given survey size and for given model parameters, the smallest detectable tensor-mode amplitude.\\
Using the expression we derived in  Eq.~(\ref{resb21}) for the non-trivial contribution to the bispectrum in the squeezed limit, one finds
\begin{equation}
\left(P^{\lambda}_{n}(K)\right)^{-1}\approx\frac{\beta^2}{\pi^2 \,15}\,K^2\,k_{\text{max}}\,,
\end{equation}
where $k_{\text{max}}$ is the maximum wave-number accessed with a given survey and the bispectrum amplitude $\beta$ in the squeezed limit has been defined, from Eq.~(\ref{resb21}), via $\|\mathcal{B}_{\text{m}}\|\equiv \beta\,P_{\gamma}^{\lambda}(K)\,P_{\zeta}(k)\frac{K}{k}$, hence $\beta= \mathcal{O}(1)\times[r_{fg}|(1-r_{fg})|]/(1+r_{fg})  $. The variance for the tensor power spectrum amplitude is then found to be 
\begin{equation}\label{aa}
\sigma_{\gamma}^{-1}=\frac{2\,\beta^2}{15\pi\sqrt{\pi}}\left(\frac{k_{\text{max}}}{k_{\text{min}}}\right)^{3/2}\,.
\end{equation}
From Eq.~(\ref{aa}) it is clear that, for an amplitude $\mathcal{A}_{\gamma}$ {close to the current observational bounds (i.e. $10^{-10}$)} to be within reach of upcoming 21-cm surveys, one would need  $\beta\gtrsim\mathcal{O}(10^3)$, {assuming $k_{\rm max}/k_{\rm min}\sim 5000$ (see e.g. \cite{Dimastrogiovanni:2014ina}).} In our configuration $r_{fg} \ll1$, therefore given the above definition for $\beta$ we conclude that the amplitude of the tensor-scalar-scalar correlation predicted by this model is too weak for detection by upcoming large scale structure surveys. Coupling $f$ directly to matter as well, via the composite metric, does allow for $r_{fg} \gg1$. However the consistency/stability requirements comprising background equations, the Higuchi bound, and the fact that it is the matter sector that ought to drive inflation, will also change the definition of $\beta$ so that a high $r_{fg}$ will not enhance the signal in the way the above expression of $\beta$ suggests.\\
\noindent A non-trivial squeezed-limit tensor-scalar-scalar bispectrum introduces also a local quadrupolar anisotropy in the scalar power spectrum which, unlike the off-diagonal correlation, is sensitive to super-horizon tensor modes \cite{Dai:2013kra}
\begin{equation}
P_\zeta(\vec{k}) |_{\gamma^{\lambda}(\vec{K})} = P_\zeta(k) \left
     [1 + \mathcal{Q}^{\lambda}_{ij}(\vec{K}) \hat{k}_i \hat{k}_{j}
     \right]\,,
\end{equation}
where 
\begin{equation}
     \mathcal{Q}^{\lambda}_{ij}(\vec{K}) \equiv
     \frac{\mathcal{B}^{}(K,k,|\vec{K}-\vec{k}|)}{P_{\gamma}(K) P_\zeta(k)} \gamma^{\lambda}_{ij}(\vec{K})\;.
\end{equation}
The variance of the quadrupole reads
\begin{equation}
     \overline{\mathcal{Q}^2} \equiv \frac{8\pi}{15}\Big\langle
     \mathcal{Q}_{ij}^{\lambda} \mathcal Q^{\lambda,ij}\Big\rangle = \frac{16}{15\pi}
     \int_{k_{\text{min}}}^{k_{\text{min}}} \, K^2 \, dK\, \left[
     \frac{\mathcal{B}_{}(K,k,|\vec{K}-\vec{k}|)}{P_{\gamma}(K) P_\zeta(k)} \right]^2 P_{\gamma}(K)\,.
\end{equation}
where  $k_{\text{min}}$ is the longest-wavelength tensor mode produced during inflation and $k_{\text{min}}$ is the smallest observable wavenumber. Replacing the results for our model, Eqs.~(\ref{resb21}), (\ref{ts}) and (\ref{cs}), one arrives at
\begin{equation}\label{qres}
 \overline{\mathcal{Q}_{}^2} \sim \frac{1}{15\pi}\left(\frac{H}{M_{P}}\right)^{2}\frac{r_{fg}^2\left(1-r_{fg}\right)^{2}}{\left(1+r_{fg}^2\right)^{2}} \left(\frac{k_{\text{min}}}{k^{\text{}}}\right)^2 \lesssim \left(\frac{H}{M_{P}}\right)^{2}\frac{r_{fg}^2\left(1-r_{fg}\right)^{2}}{\left(1+r_{fg}^2\right)^{2}}\,.
 \end{equation}
Given the current limits on the quadrupolar anisotropy, $\mathcal{Q}_{}\lesssim 10^{-2}$ \cite{Pullen:2007tu,Ando:2008zza,Groeneboom:2008fz,Hanson:2009gu,Bennett:2010jb,Pullen:2010zy,Ade:2013nlj,Kim:2013gka}, Eq.~(\ref{qres}) shows that the model allows for ample room for those to be satisfied. 

\section{Conclusions}
\label{conclusions}
Surveying the possible signatures of the inflationary particle content is crucial to make the best possible use of this cosmological window on high energy physics. The squeezed limit (and generalizations thereof) of  $N\geq3$ correlation functions is particularly sensitive a probe to the presence of ``new" physics beyond the minimal single-field inflationary scenario. Indeed, the relation between the soft limit of $N+1$-point correlators and their $N$-point counterpart may be modified as a result of the additional dynamics and reveal information about the mass and spin \cite{Kehagias:2015jha,Arkani-Hamed:2015bza,Jeong:2012df} of the extra degrees of freedom. In this work we have studied the case of the minimal inflationary scenario, GR + scalar, coupled through so-called dRGT-interactions to a (massive) spin-2 field. This is a non-trivial yet economic choice going up the particle spin ladder as we capitalize on the vast existing literature on massive spin-2 fields in the context of late-time cosmic acceleration. 
We employ a model which is ghost-free at the fully non-linear level and show how consistency relations are modified in this setup. 

The property that is most consequential for the purposes of detection turns out to be the fact that massive unitary representations of the dS group come with a lower bound on the effective mass of the graviton $m^2 \geq 2H^2$. Particles whose Compton wavelength is smaller than $1/H$ will typically decay outside the horizon. This reflects on the impact that modified consistency relations have on observables in our setup: the contribution of each mode comes only from an integrated-over-time effect. The resulting signal for the tensor-scalar-scalar correlation is too weak for detection by upcoming large scale structure surveys. One can exploit the fact that the unitarity condition is softened in FLRW backgrounds and matter may be  coupled to both spin-2 fields. However, demanding consistency of the overall setup leads to similar conclusions.

In order to rescue the massive tensor from decay, the most natural mechanism is that of a non-minimal coupling to the matter sector  \cite{Kehagias:2017cym}, where special care needs to be exerted so as not to excite ghostly degrees of freedom both at the linear and fully non-linear level. Many of these features characterize also the intriguing case of higher spin fields which will be the subject of future work.

\acknowledgments
We are delighted to thank Alex Kehagias and Toni Riotto for collaboration at the early stages of this work. We thank especially T. Riotto without whose input this paper would not have been possible. M.B. thanks Daniel Baumann, Garrett Goon, Hayden Lee and Guilherme Pimentel for helpful discussions and comments and Luigi Pilo for useful correspondence. E.D. and M.F. are grateful to Kurt Hinterbichler for fruitful conversation and especially to Claudia de Rham and Andrew J. Tolley for illuminating discussions and comments.
M.B. acknowledges support from Delta ITP consortium, a program of the Netherlands Organisation for Scientific Research (NWO) that is funded by the
Dutch Ministry of Education, Culture and Science (OCW).  E.D. acknowledges support by DOE grant DE-SC0009946. M.F. is supported in part by NSF PHY-1068380. 

\appendix
\section{Consistency relations: general considerations}
\label{crs}    
In Appendix \ref{crs} and \ref{route1} we will adopt the notation of \cite{Assassi:2012zq,Hinterbichler:2013dpa}. 
The key requirement for consistency relations to be in place between an $N+1$-function with at least one long mode and its $N$-point counterpart is the existence of gauge diffeormophisms (diff) in the action (or the equations of motion) of the physical system at hand.   
In cosmological setups (e.g. spatially-flat FLRW background) it is often said that certain gauges, for example unitary gauge, completely fix gauge invariance. This is indeed true for diffeomorphisms that vanish at spatial infinity. Crucially, the residual gauge symmetries that CRs rely upon do not vanish at infinity \cite{Weinberg:2003sw}. It can be shown \cite{Hinterbichler:2013dpa} that these transformations can be smoothly extended to configurations that do fall off at infinity and therefore corresponds to the action of adiabatic modes. The key identification is between these gauge transformations and the soft limit of transformations that fall off at infinity and are therefore physical. In the inflationary context, depending on the specific gauge diff in question, the transformation of a given observable $\mathcal{O}$ can be identified with the action of a long scalar or tensor mode on the same quantity $\mathcal{O}$.

For the CR to be non-trivial it is essential that the long mode transforms non linearly under the symmetry.  All CRs originate from the invariance of the action under specific residual space diffs:
\bea 
x^i \rightarrow x^i + {\bar \xi}^i(x) 
\label{3}
\eea
where the $\,\bar{}\,$ on top of $\xi$ indicates that the gauge mode $\xi$ must satisfy equation of motions similar to that of physical modes.  Let us consider the example of dilations, which are defined as 
\begin{align}
{\bf x}& \mapsto (1+\lambda)\bf x,	\\
\zeta &\mapsto \zeta + \lambda(1+{\bf x}\cdot \partial_{\bf x}\zeta).
\end{align}
This symmetry of the action is associated with a conserved current $J^\mu_{\rm d}$ and a corresponding charge $Q_{\rm d}=\int {\rm d}^3 x J^0_{\rm d}$, such that locally we can write  the transformation of the curvature fluctuation under dilations as
\bea
\delta_{d} \zeta = i [Q_d,\zeta]=-1 -{\bf x}\cdot \p_{\bf x} \zeta.
\eea 
 A similar relation holds for the tensor fluctuation $\gamma$ under anisotropic spatial rescaling,
\bea
\bar{\xi}_{\rm ar}^i = S_{il}\, x^l \quad{\rm with}\quad S_{il}=S_{li}\;;\;\; S_{ll}=0=\hat{q}^i S_{i l}(\vec{q})\;\;.
\label{4}
\eea
 Let us very schematically track the action of a gauge diff $s$ on a generic observable $
\langle \mathcal{O} \rangle = \langle \zeta_{1}..\zeta_{l}, \gamma_{l+1}..\gamma_m,..\sigma_N\rangle$, which we take to be the $N$-point function made up by any combinations of fields populating the Lagrangian and consider $\sigma$ as a placeholder for any type of particle content. The transformation $s$ acts according to
\bea
\delta_s \langle\mathcal{O}\rangle\Big|_{\rm connected} \propto {\bf }\mathcal D\cdot \langle\mathcal{O}\rangle\; ,
\label{ltr}
\eea 
where we have been deliberately agnostic about the rules of the $s$ transformation except for indicating by $\mathcal D\cdot $ a generic differential operator. Crucially, only the linear component of $\delta_s$ acting on the fields is relevant for connected diagrams \footnote{Note here that the action of the linear component of the $\delta_s$  transformation cannot always, strictly speaking, be put in the clean form of the RHS of Eq. (\ref{ltr}), see e.g. \cite{Hinterbichler:2013dpa}.
For example, the linear transformation of the curvature fluctuation $\zeta$ under a generic diff comprises a term proportional to the tensor mode $\gamma_{ij}$ . However, this technical consideration is in no way relevant to the discussion in this section and the simplified form is indeed in place for the specific case of anisotropic rescaling acting on two hard scalar modes.}. The very same action of the $s$ diff can be expressed in an alternative but equivalent fashion and it is the equivalence of the two that sits at the heart of consistency relations. The effect of the $\delta_s$ transformation is associated to the conserved current, and consequently to its corresponding charge so that $\delta_s \langle \mathcal{O} \rangle \sim \langle Q_s | \mathcal{O} \rangle $. At this stage we are certainly at liberty to introduce the identity operator in between
\bea
\langle Q_{\rm s} | \mathcal{O} \rangle = \sum_n \langle Q_s |n \rangle \langle n|\mathcal{O} \rangle \; ,
\eea
where the $|n\rangle$ represent mutually orthogonal independent states. Setting up this basis requires no work in the minimal inflationary scenario as naturally $\langle \zeta \gamma \rangle =0$. In the case of anisotropic rescaling, the tensor field transformation has both a linear and a non-linear component while the scalar mode only the former. It follows that $Q_{\rm ar}$ acts non-trivially only on $\gamma\;$:
\bea
\langle Q_{\rm ar} |\gamma \rangle \sim  \langle \gamma \rangle + c_{\rm number}\; ,
\eea
the first term on the RHS being zero\footnote{{The ket $| \gamma\rangle$ describes the action of the operator $\hat{\gamma}_{ij}$ on the Bunch-Davies vacuum of the theory. For a more general and detailed treatment see, for example, the work in \cite{Hinterbichler:2013dpa}.}}. With no additional field content present, it follows that:
\bea
\mathcal D\cdot \langle\mathcal{O}\rangle \propto \langle Q_{\rm ar} |\gamma\rangle \langle \gamma\,| \mathcal{O} \rangle
\eea
or, rather less schematically, the CR reads:
\bea
-\frac{1}{2}\epsilon^{\lambda}_{ij} (\hat{k}_1) \sum_{a=2}^{3} \left( k^{i}_a \frac{\p}{\p k_{a}^{j}} \langle \zeta_{{\bf k}_2} \zeta_{{\bf k}_3} \rangle^{\prime}_c \right)
= {\rm lim}_{\,{\bf k}_1 \rightarrow 0} \frac{\langle \gamma_{{\bf k}_1}^{\lambda} \zeta_{{\bf k}_2} \zeta_{{\bf k}_3}  \rangle^{\prime}_{\rm c} }{P_{\gamma}({\bf k}_1)}
\eea
where ``$c$" stands for the connected part of the diagrams and the prime symbol is a reminder that the momentum-conserving delta has been removed. Albeit admittedly very schematic, the above account is sufficient to see that one obvious route to non-standard CRs 
is the presence of an additional field whose component \textit{orthogonal} to the $\{|\gamma\rangle,|\zeta\rangle\}$ basis transforms non-linearly under anisotropic rescaling and whose ${N+1}$-point function with $\mathcal{O}$ is non-trivial \footnote{What we mean by non-trivial here is rather more than requiring a non-zero $N+1$-point function, as will be clarified below.}: 
\bea
\mathcal D\cdot \langle\mathcal{O}\rangle \propto \langle Q_{\rm ar} |\gamma \rangle \langle \gamma\,| \mathcal{O}\rangle + \langle Q_{\rm ar} |\sigma^{\,\perp} \rangle\langle \sigma^{\,\perp}\,| \mathcal{O} \rangle \;,
\label{relazione}
\eea
where $\sigma$ stands here for a generic new field. This will be precisely the case of the additional (traceless transverse component of) tensor field that will populate our non-miminal inflationary setup. Of course, there are other ways to non-standard or even broken consistency relations: one option is to have a different symmetry breaking pattern or no residual diffs altogether \cite{Endlich:2012pz,Endlich:2013dma,Bartolo:2013msa,Endlich:2013jia,Dimastrogiovanni:2014ina,Akhshik:2014bla,Cannone:2015rra,Bartolo:2015qvr,Ricciardone:2016lym}; another possibility is to have excited initial states \cite{Brahma:2013rua}. The reader might wonder how would our brief discussion change in the latter case. In the above we have been rather casual about defining the free fields eigenstates $|\zeta\rangle, |\gamma\rangle $ and the implicit vacuum wavefunction; the definition is modified in the case of non Bunch-Davies initial conditions (see \cite{Hinterbichler:2013dpa} for a more thorough discussion on this and e.g. \cite{Flauger:2013hra,Aravind:2013lra} for examples of excited initial states wavefunctionals).

\section{Alternative route to CRs breaking}
\label{route1}
In this section we shall work under the approximation that the quadratic Lagrangian (Hamiltonian) of the model, Eq.~(\ref{ggg}), can be written down as $\mathcal{L}^{(2)}_{\rm full}= \mathcal{L}^{(2)}_{0}+ \delta \mathcal{L}^{(2)}_{f-g}$ where the last piece, even though quadratic, is handled just like an interaction in the in-in formalism, with its own interaction vertex. This treatment is allowed so long as $\delta \mathcal{L}^{(2)}_{f-g}\ll\mathcal{L}^{(2)}_{0}$ and we refer the reader to \cite{Chen:2009we} for a well-studied example within the quasi-single-field inflation paradigm. This approximation is the basis for an ``alternative route" to CRs in our setup which, we stress it here, will be mostly qualitative.\\
We anticipate that treating the $f-g$ coupling in the quadratic Lagrangian as an interaction is an approximation that is no longer valid when the long mode crosses the horizon. In the approach outlined in \textit{Section \ref{calc}}, one solves for the \textit{fully} coupled equations in the quadratic Lagrangian so that all of the information on the (quadratic) $f-g$ coupling is stored in the mode-functions.

\begin{figure*}[h]
\centering
\includegraphics[scale=0.36]{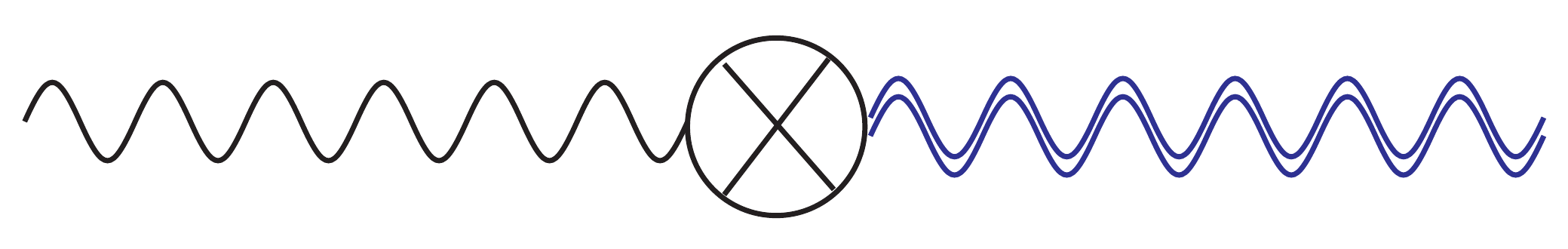}
\caption{Diagrammatic representation of the quadratic $\gamma_g$-$\gamma_f$ interaction.} 
\label{fig1}
\end{figure*}

\noi The quadratic action for the traceless transverse part of the tensors $h_f, h_g$ is schematically as follows:
\bea
\mathcal{L}^{(2)}\sim M_P^2 (\p \gamma_g)^2 + M_f^2 (\p \gamma_f)^2 + m^2 M^2 \Gamma(H/H_f)( \gamma_g^2  + \gamma_g \gamma_f+ \gamma_f^2) 
\label{schematic}
\eea
where $\Gamma$ is a specific function of the $\beta_n$s (one can think of them as order one) and the $H/H_f$ ratio. The latter is not too relevant here as the $M_f/M_P$ ratio will play the main role. We want to show that the $\gamma_g-\gamma_f$ mixing can be treated as a perturbation on top of the remaining free Lagrangian. To this end, we recall that the usual normalization factor for the wavefunctions is $\gamma_g\propto \frac{H}{M_P}$ and $ \gamma_f\propto \frac{H_f}{M_f}$, so that the Einstein-Hilbert contributions to the quadratic Lagrangian go rispectively like $H^4, H_f^4$ and we take the former as being of the same order or smaller than the latter. It follows that the condition for each term proportional to $M^2$ in Eq. (\ref{schematic}) to be smaller than the E-H terms is, in order:
\bea
\frac{m^2}{H^2}\frac{M^2}{M_P^2}\Gamma \ll 1 \,, \quad \frac{m^2}{H^2}\frac{H_f}{H}\frac{M^2}{M_f M_P}\Gamma\ll 1 \,, \quad \frac{m^2}{H^2}\frac{H_f^2}{H^2}\frac{M^2}{M_f^2 }\Gamma\ll 1\, ,
\label{routea}
\eea
the most relevant being of course the second inequality because it stems from requiring that the $\gamma_g-\gamma_f$ mixing term can be treated as an interaction. It is important to check at this stage the compatibility of the second inequality\footnote{As we shall see, the first one is a byproduct of the second one in our parameter space and we will not need the third inequality to be satisfied at all.} with the Higuchi bound as well as the Friedman equations. For convenience, we report the Higuchi bound as in Eq. (\ref{higuchi}) and assume the term regulated by $\beta_1$ to be the leading one. The bound to be compared with the conditions in Eq.~(\ref{routea}) reads:
\bea
\frac{1}{4}\frac{m^2}{H^2} \frac{H_f}{H} \frac{M^2}{M_f^2}\geq1\; .
\eea
Upon inspection, it is immediate that the first two requirements in Eq.~(\ref{routea}) are satisfied if the hierarchy $M_f\ll M_P$ is in place \footnote{We must stress here that, at this stage, this is not a necessary assumption. However, the combination of Higuchi bound,  the requirement on the Friedman equation to be dominated by matter driving inflation, and having a slowly varying $\xi$, leads to $M_f\ll M_P$. See \cite{DeFelice:2014nja,Fasiello:2015csa} for further details.}; we shall assume it for the remaining parts of the manuscript\footnote{The third requirement in Eq.~(\ref{routea}) is not necessarily satisfied: the ratio can be order one and translates into a massive term for $\gamma_f$ that must be contained within $\mathcal{L}^{(2)}_0$ and is not a part of $\delta \mathcal{L}^{(2)}$ . The same ratio appears e.g. in Eq.~(\ref{boundfrhf}).}. This in particular implies that the $\gamma_g-\gamma_f$ mixing can be treated as an interaction (see Fig.~\ref{fig1}) without conflicting with the all-important unitarity bound. For consistency, we shall also require that the leading part in the RHS of Eq.~(\ref{frhg}) is also due to the inflationary potential and not the $m^2$-proportional part. To that end, and again focussing the attention on the $\propto \beta_1$ contribution, we ought to require that:
\bea
\frac{m^2}{H^2} \frac{H}{H_f} \frac{M^2}{M_P^2}\ll1\; ,
\eea
which is compatible with all of the above. Finally, Eq.~(\ref{frhf}) requires
\bea
\frac{m^2}{H^2} \frac{H_f}{H} \frac{M^2}{M_f^2}\sim 1\; ,
\label{boundfrhf}
\eea
also consistent. Reassured by these checks, let us turn to consistency relations in this approximation scheme.

\noi We want to provide evidence that tensor CRs in our setup take a non standard form.  We have reviewed in \textit{Section \ref{crs}} the origin of the CRs relevant for observables such as the tensor-scalar-scalar three-point function.  We are interested in the action of anisotropic rescaling on long tensor modes defined in Eq. \eqref{4}.
\noindent Recall now from Eq.~(\ref{relazione}) that a non-standard CR demands at least two terms in the RHS of
\bea
\langle Q_{\rm ar}| \zeta\zeta \rangle = \sum_n \langle Q_{\rm ar}  |n \rangle \langle n|\zeta\zeta \rangle \; .
\label{8}
\eea 
\noi 
If at least two fields (or rather two states) $\{|1 \rangle \equiv |\gamma_g  \rangle, |2 \rangle,\dots\}$, once orthonormalized, transform non-linearly under $Q_{\rm ar}$ then we will have a non-standard CR:
\bea
\langle Q_{\rm ar}  |\gamma_g \rangle \langle \gamma_g|\zeta\zeta \rangle + \langle Q_{\rm ar}  |2 \rangle \langle 2|\zeta\zeta \rangle +... \sim  \langle[Q_{\rm ar}, \zeta\zeta]\rangle \sim \mathcal D \langle \zeta\zeta\rangle
\label{9}
\eea
In standard single-field inflation the only field non-linearly transforming under an anisotropic rescaling is $\gamma_g$, hence the standard CR.  In the setup of Eq.~(\ref{model}), can $\gamma_f$, the traceless and transverse part of the fluctuation of the $f$ metric, suitably orthogonalized, serve as a non zero contribution from the $| 2 \rangle$ ``state"? To answer this question in the affirmative we ought to show that 

\bea
{0\not=}\langle Q_{\rm ar}  |\gamma^{\perp}_f \rangle \langle \gamma^{\perp}_f|\zeta\zeta \rangle \not=\langle Q_{\rm ar}  |\gamma^{}_g \rangle \langle \gamma^{}_g|\zeta\zeta \rangle \; ,
\eea
the last condition guaranteeing that the action of the two long modes $\gamma_f$ and $\gamma_g$ is not identical. Note also that we have orthogonalized $\gamma_f$ w.r.t. $\gamma_g$ in the standard way:
\bea
|\gamma_f^{\perp} \rangle = N\Big[|\gamma_f \rangle- |\gamma_g \rangle \langle \gamma_g | \gamma_f\rangle    \Big], 
\eea 
with $N$ a normalization factor. The RHS of Eq.~(\ref{9}) can now be written more explicitly as
\bea
&&\sum _n\langle Q_{\rm ar}  |n \rangle \langle n|\zeta\zeta \rangle= \langle Q_{\rm ar}  |\gamma^{}_g \rangle \langle \gamma^{}_g|\zeta\zeta \rangle+ \langle Q_{\rm ar}  |\gamma^{\perp}_f \rangle \langle \gamma^{\perp}_f|\zeta\zeta \rangle = \nonumber \\
&&  c_g \langle \gamma^{}_g|\zeta\zeta \rangle+ \langle Q_{\rm ar}  | N \Big[|\gamma_f \rangle- |\gamma_g \rangle \langle \gamma_g | \gamma_f\rangle    \Big]  \cdot N^{*}\Big[\langle \gamma_f|  - \langle \gamma_f | \gamma_g\rangle \langle \gamma_g |     \Big]  |\zeta\zeta \rangle = 
\nonumber \\
&& c_g \langle \gamma^{}_g|\zeta\zeta \rangle+ |N|^2\Big[   \left( { c_f - c_g I_{g\,f} } \right) \cdot \left(\,\langle \gamma_f| \zeta \zeta \rangle  - I_{g\,f}^{*} \langle \gamma_g| \zeta \zeta \rangle\, \right) \Big],
\label{long}
\eea
where we have assumed for simplicity that  $| \gamma\rangle$  has norm one. We have also defined the  quantities $c_g\equiv \langle Q_{\rm ar}| \gamma_g\rangle$ , $c_f\equiv \langle Q_{\rm ar}| \gamma_f\rangle$ , and $I_{g\, f} \equiv \langle \gamma_g|\gamma_f \rangle$ . Note that the latter stems precisely from the interaction in Fig.~\ref{fig1}. Non-standard CRs now correspond to both factors between square brackets in the last line of Eq.~(\ref{long}) being non-zero. In order to show that is indeed the case, let us write down the gauge transformation properties of $\gamma_f, \gamma_g$ . These, in our language, are  stored into $c_f, c_g$. For the gauge parameter, $\xi^{\mu}$, the transformation law reads: 
\bea
\begin{split}
& \delta \gamma_{g \, \mu \nu } = a^{-2} \left(\xi^\alpha \p_\alpha  {\bar g}_{ \,
  \mu \nu} +  \bar g_{ \, \alpha \nu}  \, \p_\mu
  \xi^\alpha + \bar g_{ \, \mu \alpha}  \, \p_\nu \xi^\alpha \right)
\, , \\[.2cm]
&\delta \gamma_{f \, \mu \nu } = b^{-2} \left(\xi^\alpha \p_\alpha \, {\bar
  f}_{ \,\mu \nu } +  \bar f_{ \, \alpha \nu}  \, \p_\mu
  \xi^\alpha + \bar f_{ \, \mu \alpha}  \, \p_\nu \xi^\alpha \right) \, .
\end{split}
\eea

\begin{figure*}[h]
\centering
\includegraphics[scale=0.6]{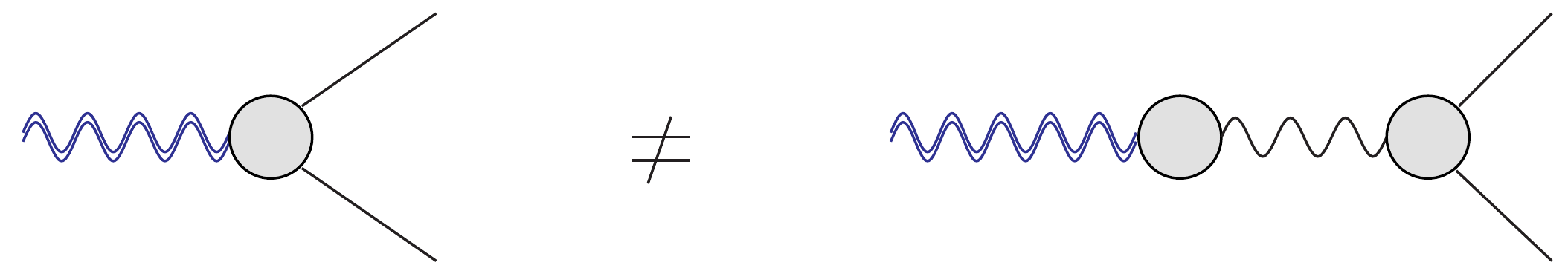}
\caption{Diagrammatic representation non standard CRs condition.} 
\label{fig2}
\end{figure*}

\begin{figure*}[h]
\centering
\includegraphics[scale=0.60]{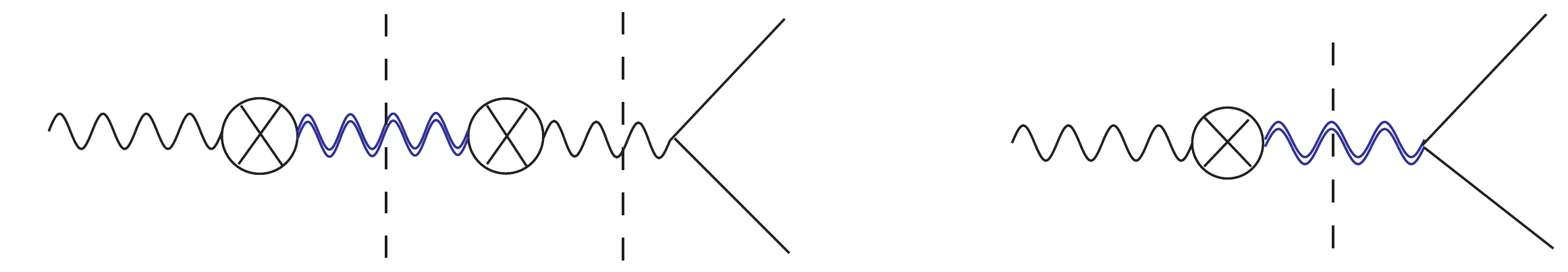}
\caption{Left: contribution to $\langle \gamma_g \zeta \zeta \rangle$ consisting of two ``quadratic" $\gamma_g-\gamma_f$ vertices and the usual tree-level tensor-scalar-scalar interaction. Right: A $\gamma_g-\gamma_f$ vertex and a three-vertex which is there only if $\gamma_f$ couples directly with $\zeta$.} 
\label{fig3}
\end{figure*}

\noi  The reasoning proceeds as follows. The gauge transformation of $\gamma_f, \gamma_g$ are the same except for a factor of $b/a$, the ratio of the two metrics scale factors, a quantity also equal to $H/H_f$ in the branch of the theory we are considering. From here, one should in principle solve the constraint equations and derive the transformation law specifically for the symmetric traceless transverse part of  $\gamma_f, \gamma_g$.  On the other hand, what matters for us is that the following term from Eq.~(\ref{long}) is non zero: $c_f - c_g I_{g\,f}\not=0$. It is immediate to see, for example from the function $\Gamma$ in Eq.~(\ref{schematic}), that the quadratic interaction term $I_{g\, f}$ contains different powers of $b/a=H/H_f$ and is also a function of $\beta_1,\beta_2,\beta_3$ so that $c_f - c_g I_{g\,f}$ does not in general vanish.\\
What remains to be proven now is that $\langle \gamma_f| \zeta \zeta \rangle  - I_{g\,f}^{*} \langle \gamma_g| \zeta \zeta \rangle\not=0$ which is, written differently, $\langle \gamma_f| \zeta \zeta \rangle  \not= \langle \gamma_f |\gamma_g\rangle \langle \gamma_g| \zeta \zeta \rangle$. Perhaps the best way to prove this last part is diagrammatically, as shown in Fig.~\ref{fig2}, where the grey colored circles stand for  generic interactions. It will be enough to find an internal\footnote{In this approximation $\gamma_f$ is massive and will therefore decay at late times.} diagram corresponding to $\langle \gamma_f| \zeta \zeta \rangle$  that cannot be expressed as $\langle \gamma_f |\gamma_g\rangle \langle \gamma_g| \zeta \zeta \rangle$.
\noi  Let us give a few examples of interactions that do and do not lead to a modified CR. On the left side of Fig.~\ref{fig3} one can see a diagram that can be drawn as in the RHS of Fig.~{\ref{fig2}} with dashed vertical lines just to the left of where the gray vertices would go.

\noi On the other hand, the diagram on the right side of Fig.~\ref{fig3} cannot be put in that form and leads therefore to a non-standard consistency relation. 

\begin{figure*}[h]
\centering
\includegraphics[scale=0.50]{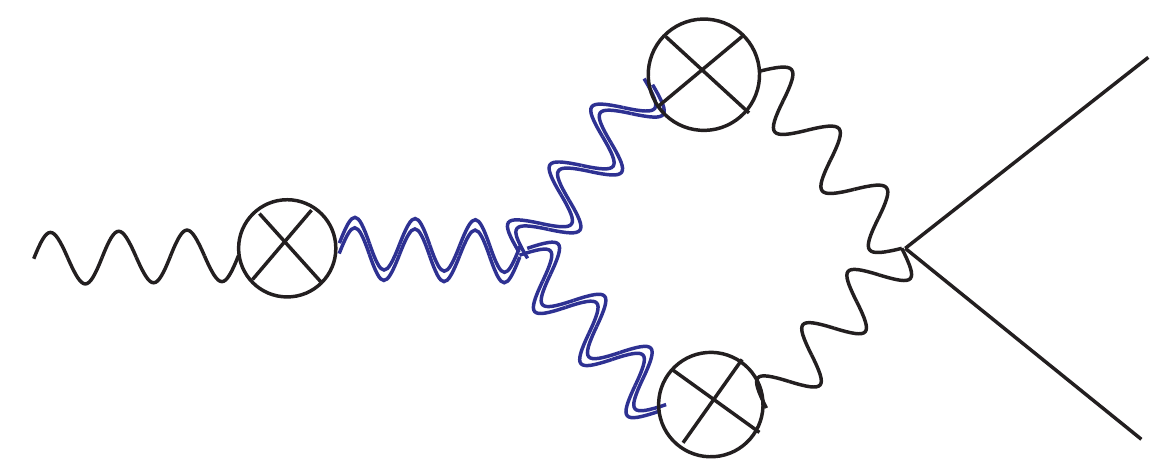}
\caption{A loop-diagram contribution to $\langle \gamma_g \zeta \zeta \rangle$. This diagram will be ``Planck" suppressed by the  $H/M_p$ and $H_f/M_f$ in the wavefunctions normalizations.} 
\label{fig4}
\end{figure*}

\noi However, it must be noted that this requires a direct coupling between $\gamma_f$ and $\zeta$. The diagram in Fig.~\ref{fig4} is instead a loop effect and delivers a non-standard CR regardless of the coupling of $\gamma_f$. We consider this last one as the clearest example of a modified consistency relation among those we present in this  \textit{Section \ref{route1}}. As anticipated, we will not discuss in detail the value of the bispectrum generated by these diagrams, we postpone a quantitative analysis of modified CRs to \textit{Section \ref{route2}}. One reason, as we shall see shortly, is that the approximation scheme justifying the use of $\delta\mathcal{L}^{(2)}$ , or in other words the diagram in Fig.~\ref{fig1}, breaks down at the horizon. Nevertheless, we find it worthwhile to present a diagrammatic proof of CRs ``breaking" in this setup. The existence of a quadratic interaction lends itself nicely to such proof, it shows very clearly why CRs are indeed broken, and is very reminiscent of the one that applies to, among others, quasi-single-field inflation models.   

\noi Let us briefly show how the approximation breaks down at the horizon. The reason is quite simple, the quadratic interaction term is proportional to $m^2 a(\tau)^2 h^2$. In a quasi-de Sitter phase it is true that $a^2\simeq -\frac{1}{H^2 \tau^2}$. At late times (starting at horizon exit) this term becomes of the same order and eventually much larger than the Einstein-Hilbert terms we have placed in $\mathcal{L}^{(2)}_0$, thereby invalidating the approximation. Considering that most of the contribution to cosmological correlation functions is known to originate at and after horizon exit it is advisable to seek another scheme whose validity extends further than this ``alternative route". We provided such a treatment in \textit{Section \ref{route2}}.


\begin{thebibliography}{99}


\bibitem{Kehagias:2015jha} 
  A.~Kehagias and A.~Riotto,
  Fortsch.\ Phys.\  {\bf 63}, 531 (2015)
     [\href{https://arxiv.org/abs/1501.03515}{{\blu{arXiv:1501.03515}}}].
     
     
\bibitem{Arkani-Hamed:2015bza} 
  N.~Arkani-Hamed and J.~Maldacena,
      [\href{https://arxiv.org/abs/1503.08043}{{\blu{arXiv:1503.08043}}}].

  
\bibitem{Dimastrogiovanni:2015pla} 
  E.~Dimastrogiovanni, M.~Fasiello and M.~Kamionkowski,
  JCAP {\bf 1602}, 017 (2016)
     [\href{https://arxiv.org/abs/1504.05993}{{\blu{arXiv:1504.05993}}}].

\bibitem{Lee:2016vti} 
  H.~Lee, D.~Baumann and G.~L.~Pimentel,
  JHEP {\bf 1612}, 040 (2016)
       [\href{https://arxiv.org/abs/1607.03735}{{\blu{arXiv:1607.03735}}}].
  
\bibitem{Baumann:2014nda} 
  D.~Baumann and L.~McAllister,
         [\href{https://arxiv.org/abs/1404.2601}{{\blu{arXiv:1404.2601}}}].



\bibitem{Achucarro:2012sm}
  A.~Achucarro, J.~O.~Gong, S.~Hardeman, G.~A.~Palma and S.~P.~Patil,
  JHEP {\bf 1205} (2012) 066
           [\href{https://arxiv.org/abs/1201.6342}{{\blu{arXiv:1201.6342}}}].



\bibitem{Burgess:2012dz}
  C.~P.~Burgess, M.~W.~Horbatsch and S.~P.~Patil,
  JHEP {\bf 1301} (2013) 133
           [\href{https://arxiv.org/abs/1209.5701}{{\blu{arXiv:1209.5701}}}].



\bibitem{Silverstein:2017zfk} 
  E.~Silverstein,
         [\href{https://arxiv.org/abs/1706.02790}{{\blu{arXiv:1706.02790}}}].


\bibitem{Chen:2009we} 
  X.~Chen and Y.~Wang,
  Phys.\ Rev.\ D {\bf 81}, 063511 (2010)
   [\href{https://arxiv.org/abs/0909.0496}{{\blu{arXiv:0909.0496}}}].

\bibitem{Chen:2009zp} 
  X.~Chen and Y.~Wang,
  JCAP {\bf 1004}, 027 (2010)
   [\href{https://arxiv.org/abs/0911.3380}{{\blu{arXiv:0911.3380}}}].
  
\bibitem{Craig:2014rta} 
  N.~Craig and D.~Green,
  JHEP {\bf 1407}, 102 (2014)
     [\href{https://arxiv.org/abs/1403.7193}{{\blu{arXiv:1403.7193}}}].
  
\bibitem{Noumi:2012vr} 
  T.~Noumi, M.~Yamaguchi and D.~Yokoyama,
  JHEP {\bf 1306}, 051 (2013)
   [\href{https://arxiv.org/abs/1211.1624}{{\blu{arXiv:1211.1624}}}].

\bibitem{Sefusatti:2012ye} 
  E.~Sefusatti, J.~R.~Fergusson, X.~Chen and E.~P.~S.~Shellard,
  JCAP {\bf 1208}, 033 (2012)
     [\href{https://arxiv.org/abs/1204.6318}{{\blu{arXiv:1204.6318}}}].
  
\bibitem{McAllister:2012am} 
  L.~McAllister, S.~Renaux-Petel and G.~Xu,
  JCAP {\bf 1210}, 046 (2012)
       [\href{https://arxiv.org/abs/1207.0317}{{\blu{arXiv:1207.0317}}}].

\bibitem{Baumann:2011nk} 
  D.~Baumann and D.~Green,
  Phys.\ Rev.\ D {\bf 85}, 103520 (2012)
    [\href{https://arxiv.org/abs/1109.0292}{{\blu{arXiv:1109.0292}}}].




\bibitem{Higuchi:1986py} 
  A.~Higuchi,
  Nucl.\ Phys.\ B {\bf 282}, 397 (1987).
  
\bibitem{Blas:2007zza} 
  D.~Blas,
  Int.\ J.\ Theor.\ Phys.\  {\bf 46}, 2258 (2007).
  
    
\bibitem{Comelli:2012db} 
  D.~Comelli, M.~Crisostomi and L.~Pilo,
  JHEP {\bf 1206}, 085 (2012)
    [\href{https://arxiv.org/abs/1202.1986}{{\blu{arXiv:1202.1986}}}].
  
\bibitem{Fasiello:2012rw} 
  M.~Fasiello and A.~J.~Tolley,
  JCAP {\bf 1211}, 035 (2012)
   [\href{https://arxiv.org/abs/1206.3852}{{\blu{arXiv:1206.3852}}}].


\bibitem{Fasiello:2013woa} 
  M.~Fasiello and A.~J.~Tolley,
  JCAP {\bf 1312}, 002 (2013)
   [\href{https://arxiv.org/abs/1308.1647}{{\blu{arXiv:1308.1647}}}].
  
  
\bibitem{Ade:2015ava} 
  P.~A.~R.~Ade {\it et al.} [Planck Collaboration],
  Astron.\ Astrophys.\  {\bf 594}, A17 (2016)
 [\href{https://arxiv.org/abs/1502.01592}{{\blu{arXiv:1502.01592}}}].
 


\bibitem{Maldacena:2002vr} 
  J.~M.~Maldacena,
  JHEP {\bf 0305}, 013 (2003)
      [\href{https://arxiv.org/abs/astro-ph/0210603}{{\blu{arXiv:0210603}}}].\\
  
\bibitem{Creminelli:2004yq} 
  P.~Creminelli and M.~Zaldarriaga,
  JCAP {\bf 0410}, 006 (2004)
        [\href{https://arxiv.org/abs/astro-ph/0407059}{{\blu{arXiv:0407059}}}].\\
  
\bibitem{Assassi:2012zq} 
  V.~Assassi, D.~Baumann and D.~Green,
  JCAP {\bf 1211}, 047 (2012)
    [\href{https://arxiv.org/abs/1204.4207}{{\blu{arXiv:1204.4207}}}].\\
  
  
\bibitem{Hinterbichler:2013dpa} 
  K.~Hinterbichler, L.~Hui and J.~Khoury,
  JCAP {\bf 1401}, 039 (2014)
      [\href{https://arxiv.org/abs/1304.5527}{{\blu{arXiv:1304.5527}}}].\\

\bibitem{Tanaka:2017nff} 
  T.~Tanaka and Y.~Urakawa,
     [\href{https://arxiv.org/abs/1707.05485}{{\blu{arXiv:1707.05485}}}].\\




\bibitem{Kehagias:2013yd} 
  A.~Kehagias and A.~Riotto,
  Nucl.\ Phys.\ B {\bf 873}, 514 (2013)
         [\href{https://arxiv.org/abs/1302.0130}{{\blu{arXiv:1302.0130}}}].



\bibitem{Peloso:2013zw} 
  M.~Peloso and M.~Pietroni,
  JCAP {\bf 1305}, 031 (2013)
         [\href{https://arxiv.org/abs/1302.0223}{{\blu{arXiv:1302.0223}}}].

  
\bibitem{Kehagias:2013rpa} 
  A.~Kehagias, J.~Nore\~{n}a, H.~Perrier and A.~Riotto,
  Nucl.\ Phys.\ B {\bf 883}, 83 (2014)
  [\href{https://arxiv.org/abs/1311.0786}{{\blu{arXiv:1311.0786}}}].
  
  
\bibitem{Creminelli:2013mca} 
  P.~Creminelli, J.~Nore\~{n}a, M.~Simonovi\'c and F.~Vernizzi,
  JCAP {\bf 1312}, 025 (2013)
         [\href{https://arxiv.org/abs/1309.3557}{{\blu{arXiv:1309.3557}}}].

  
\bibitem{Kehagias:2013paa}
  A.~Kehagias, H.~Perrier and A.~Riotto,
  Mod.\ Phys.\ Lett.\ A {\bf 29} (2014) 1450152
   [\href{https://arxiv.org/abs/1311.5524}{{\blu{arXiv:1311.5524}}}].
  
\bibitem{Creminelli:2013nua} 
  P.~Creminelli, J.~Gleyzes, L.~Hui, M.~Simonovi\'c and F.~Vernizzi,
  JCAP {\bf 1406}, 009 (2014)
   [\href{https://arxiv.org/abs/1312.6074}{{\blu{arXiv:1312.6074}}}].
 
 
 
  
\bibitem{Horn:2014rta} 
  B.~Horn, L.~Hui and X.~Xiao,
  JCAP {\bf 1409}, no. 09, 044 (2014)
      [\href{https://arxiv.org/abs/1406.0842}{{\blu{arXiv:1406.0842}}}].
  
\bibitem{Kehagias:2015tda}
  A.~Kehagias, A.~M.~Dizgah, J.~Nore\~{n}a, H.~Perrier and A.~Riotto,
  JCAP {\bf 1508} (2015) no.08,  018
       [\href{https://arxiv.org/abs/1503.04467}{{\blu{arXiv:1503.04467}}}].
  
  
\bibitem{DiDio:2016gpd} 
  E.~Di Dio, H.~Perrier, R.~Durrer, G.~Marozzi, A.~M.~Dizgah, J.~Nore\~{n}a and A.~Riotto,
  JCAP {\bf 1703}, no. 03, 006 (2017)
 [\href{https://arxiv.org/abs/1611.03720}{{\blu{arXiv:1611.03720}}}].
  
    
\bibitem{Fasiello:2016yvr} 
  M.~Fasiello and Z.~Vlah,
         [\href{https://arxiv.org/abs/1611.00542}{{\blu{arXiv:1611.00542}}}].
  

  
\bibitem{Bonga:2015urq} 
  B.~Bonga, S.~Brahma, A.~S.~Deutsch and S.~Shandera,
  JCAP {\bf 1605}, no. 05, 018 (2016)
     [\href{https://arxiv.org/abs/1512.05365}{{\blu{arXiv:1512.05365}}}].



\bibitem{deRham:2010ik} 
  C.~de Rham and G.~Gabadadze,
  Phys.\ Rev.\ D {\bf 82}, 044020 (2010)
       [\href{https://arxiv.org/abs/1007.0443}{{\blu{arXiv:1007.0443}}}].

\bibitem{deRham:2010kj} 
  C.~de Rham, G.~Gabadadze and A.~J.~Tolley,
  Phys.\ Rev.\ Lett.\  {\bf 106}, 231101 (2011)
       [\href{https://arxiv.org/abs/1011.1232}{{\blu{arXiv:1011.1232}}}].
  
\bibitem{Hassan:2011tf} 
  S.~F.~Hassan, R.~A.~Rosen and A.~Schmidt-May,
  JHEP {\bf 1202}, 026 (2012)
       [\href{https://arxiv.org/abs/1109.3230}{{\blu{arXiv:1109.3230}}}].








\bibitem{Weinberg:2003sw} 
  S.~Weinberg,
  Phys.\ Rev.\ D {\bf 67}, 123504 (2003)
   [\href{https://arxiv.org/abs/astro-ph/0302326}{{\blu{arXiv:0302326}}}].
  


\bibitem{Endlich:2012pz} 
  S.~Endlich, A.~Nicolis and J.~Wang,
  JCAP {\bf 1310}, 011 (2013)
  [\href{https://arxiv.org/abs/1210.0569}{{\blu{arXiv:1210.0569}}}].

\bibitem{Endlich:2013dma} 
  S.~Endlich and A.~Nicolis,
    [\href{https://arxiv.org/abs/1303.3289}{{\blu{arXiv:1303.3289}}}].

  
\bibitem{Bartolo:2013msa} 
  N.~Bartolo, S.~Matarrese, M.~Peloso and A.~Ricciardone,
  JCAP {\bf 1308}, 022 (2013)
      [\href{https://arxiv.org/abs/1306.4160}{{\blu{arXiv:1306.4160}}}].
  
  
\bibitem{Endlich:2013jia} 
  S.~Endlich, B.~Horn, A.~Nicolis and J.~Wang,
  Phys.\ Rev.\ D {\bf 90}, no. 6, 063506 (2014)
      [\href{https://arxiv.org/abs/1307.8114}{{\blu{arXiv:1307.8114}}}].


\bibitem{Dimastrogiovanni:2014ina} 
  E.~Dimastrogiovanni, M.~Fasiello, D.~Jeong and M.~Kamionkowski,
  JCAP {\bf 1412}, 050 (2014)
  [\href{https://arxiv.org/abs/1407.8204}{{\blu{arXiv:1407.8204}}}].



\bibitem{Akhshik:2014bla} 
  M.~Akhshik,
  JCAP {\bf 1505}, no. 05, 043 (2015)
      [\href{https://arxiv.org/abs/1409.3004}{{\blu{arXiv:1409.3004}}}].


\bibitem{Cannone:2015rra} 
  D.~Cannone, J.~O.~Gong and G.~Tasinato,
  JCAP {\bf 1508}, no. 08, 003 (2015)
        [\href{https://arxiv.org/abs/1505.05773}{{\blu{arXiv:1505.05773}}}].

\bibitem{Bartolo:2015qvr} 
  N.~Bartolo, D.~Cannone, A.~Ricciardone and G.~Tasinato,
  JCAP {\bf 1603}, no. 03, 044 (2016)
      [\href{https://arxiv.org/abs/1511.07414}{{\blu{arXiv:1511.07414}}}].
  
\bibitem{Ricciardone:2016lym}
  A.~Ricciardone and G.~Tasinato,
    [\href{https://arxiv.org/abs/1611.04516}{{\blu{arXiv:1611.04516}}}].





\bibitem{Brahma:2013rua} 
  S.~Brahma, E.~Nelson and S.~Shandera,
  Phys.\ Rev.\ D {\bf 89}, no. 2, 023507 (2014)
  [\href{https://arxiv.org/abs/1310.0471}{{\blu{arXiv:1310.0471}}}].









  
\bibitem{Flauger:2013hra} 
  R.~Flauger, D.~Green and R.~A.~Porto,
  JCAP {\bf 1308}, 032 (2013)
         [\href{https://arxiv.org/abs/1303.1430}{{\blu{arXiv:1303.1430}}}].
  
\bibitem{Aravind:2013lra} 
  A.~Aravind, D.~Lorshbough and S.~Paban,
  JHEP {\bf 1307}, 076 (2013)
     [\href{https://arxiv.org/abs/1303.1440}{{\blu{arXiv:1303.1440}}}].
  

  


\bibitem{Boulanger:2000rq} 
  N.~Boulanger, T.~Damour, L.~Gualtieri and M.~Henneaux,
  Nucl.\ Phys.\ B {\bf 597}, 127 (2001)
        [\href{https://arxiv.org/abs/hep-th/0007220}{{\blu{arXiv:0007220}}}].\\

\bibitem{Riess:1998cb} 
  A.~G.~Riess {\it et al.} [Supernova Search Team],
  Astron.\ J.\  {\bf 116}, 1009 (1998)
          [\href{https://arxiv.org/abs/astro-ph/9805201}{{\blu{arXiv:9805201}}}].\\
  
\bibitem{Schmidt:1998ys}
  B.~P.~Schmidt {\it et al.} [Supernova Search Team],
  Astrophys.\ J.\  {\bf 507} (1998) 46
   [\href{https://arxiv.org/abs/astro-ph/9805200}{{\blu{arXiv:9805200}}}].

\bibitem{Perlmutter:1998hx}
  S.~Perlmutter {\it et al.} [Supernova Cosmology Project Collaboration],
  Bull.\ Am.\ Astron.\ Soc.\  {\bf 29} (1997) 1351
  [astro-ph/9812473].
   [\href{https://arxiv.org/abs/astro-ph/9812473}{{\blu{arXiv:9805200}}}].

  
  
\bibitem{Vegh:2013sk} 
  D.~Vegh,
       [\href{https://arxiv.org/abs/1301.0537}{{\blu{arXiv:1301.0537}}}].
  
\bibitem{Blake:2013owa} 
  M.~Blake, D.~Tong and D.~Vegh,
  Phys.\ Rev.\ Lett.\  {\bf 112}, no. 7, 071602 (2014)
     [\href{https://arxiv.org/abs/1310.3832}{{\blu{arXiv:1310.3832}}}].
  
\bibitem{Vainshtein:1972sx} 
  A.~I.~Vainshtein,
  Phys.\ Lett.\  {\bf 39B}, 393 (1972).
  
\bibitem{Babichev:2013usa} 
  E.~Babichev and C.~Deffayet,
  Class.\ Quant.\ Grav.\  {\bf 30}, 184001 (2013)
       [\href{https://arxiv.org/abs/1304.7240}{{\blu{arXiv:1304.7240}}}].
  
\bibitem{DeFelice:2014nja} 
  A.~De Felice, A.~E.~Gumrukcuoglu, S.~Mukohyama, N.~Tanahashi and T.~Tanaka,
  JCAP {\bf 1406}, 037 (2014)
    [\href{https://arxiv.org/abs/1404.0008}{{\blu{arXiv:1404.0008}}}].\\
  
  
\bibitem{Lagos:2014lca} 
  M.~Lagos and P.~G.~Ferreira,
  JCAP {\bf 1412}, 026 (2014)
        [\href{https://arxiv.org/abs/1410.0207}{{\blu{arXiv:1410.0207}}}].\\
  
\bibitem{Johnson:2015tfa} 
  M.~Johnson and A.~Terrana,
  Phys.\ Rev.\ D {\bf 92}, no. 4, 044001 (2015)
        [\href{https://arxiv.org/abs/1503.05560}{{\blu{arXiv:1503.05560}}}].\\
  
\bibitem{Akrami:2015qga} 
  Y.~Akrami, S.~F.~Hassan, F.~Könnig, A.~Schmidt-May and A.~R.~Solomon,
  Phys.\ Lett.\ B {\bf 748}, 37 (2015)
        [\href{https://arxiv.org/abs/1503.07521}{{\blu{arXiv:1503.07521}}}].\\
  
  
  
\bibitem{Cusin:2015pya} 
  G.~Cusin, R.~Durrer, P.~Guarato and M.~Motta,
  JCAP {\bf 1509}, no. 09, 043 (2015)
      [\href{https://arxiv.org/abs/1505.01091}{{\blu{arXiv:1505.01091}}}].\\
  
  
  
\bibitem{Brax:2017hxh} 
  P.~Brax, A.~C.~Davis and J.~Noller,
        [\href{https://arxiv.org/abs/1703.08016}{{\blu{arXiv:1703.08016}}}].\\
  
\bibitem{deRham:2014zqa} 
  C.~de Rham,
  Living Rev.\ Rel.\  {\bf 17}, 7 (2014)
 [\href{https://arxiv.org/abs/1401.4173}{{\blu{arXiv:1401.4173}}}].
  
  
\bibitem{Kehagias:2017rpe} 
  A.~Kehagias and A.~Riotto,
  Fortsch.\ Phys.\  {\bf 65}, no. 5, 1700023 (2017)
          [\href{https://arxiv.org/abs/1701.05462}{{\blu{arXiv:1701.05462}}}].\\
  
\bibitem{deRham:2014naa} 
  C.~de Rham, L.~Heisenberg and R.~H.~Ribeiro,
  Class.\ Quant.\ Grav.\  {\bf 32}, 035022 (2015)
         [\href{https://arxiv.org/abs/1408.1678}{{\blu{arXiv:1408.1678}}}].\\
  
  
\bibitem{Fasiello:2015csa} 
  M.~Fasiello and R.~H.~Ribeiro,
  JCAP {\bf 1507}, no. 07, 027 (2015)
      [\href{https://arxiv.org/abs/1505.00404}{{\blu{arXiv:1505.00404}}}].






%

\bibitem{Jeong:2012df} 
  D.~Jeong and M.~Kamionkowski,
  Phys.\ Rev.\ Lett.\  {\bf 108}, 251301 (2012)
    [\href{https://arxiv.org/abs/1203.0302}{{\blu{arXiv:1203.0302}}}].

\bibitem{Dai:2013ikl} 
  L.~Dai, D.~Jeong and M.~Kamionkowski,
  Phys.\ Rev.\ D {\bf 87}, no. 10, 103006 (2013)
   [\href{https://arxiv.org/abs/1302.1868}{{\blu{arXiv:1302.1868}}}].


%

\bibitem{Dai:2013kra} 
  L.~Dai, D.~Jeong and M.~Kamionkowski,
  Phys.\ Rev.\ D {\bf 88}, no. 4, 043507 (2013)
   [\href{https://arxiv.org/abs/1306.3985}{{\blu{arXiv:1306.3985}}}].\\
  
  
  
  
  %
  
\bibitem{Pullen:2007tu} 
  A.~R.~Pullen and M.~Kamionkowski,
  Phys.\ Rev.\ D {\bf 76}, 103529 (2007)
   [\href{https://arxiv.org/abs/0709.1144}{{\blu{arXiv:0709.1144}}}].
  
  
\bibitem{Ando:2008zza} 
  S.~Ando and M.~Kamionkowski,
  Phys.\ Rev.\ Lett.\  {\bf 100}, 071301 (2008)
   [\href{https://arxiv.org/abs/0711.0779}{{\blu{arXiv:0711.0779}}}].


\bibitem{Groeneboom:2008fz} 
  N.~E.~Groeneboom and H.~K.~Eriksen,
  Astrophys.\ J.\  {\bf 690}, 1807 (2009)
  [\href{https://arxiv.org/abs/0807.2242}{{\blu{arXiv:0807.2242}}}].

\bibitem{Hanson:2009gu} 
  D.~Hanson and A.~Lewis,
  Phys.\ Rev.\ D {\bf 80}, 063004 (2009)
    [\href{https://arxiv.org/abs/0908.0963}{{\blu{arXiv:0908.0963}}}].
  
  
\bibitem{Bennett:2010jb} 
  C.~L.~Bennett {\it et al.},
  Astrophys.\ J.\ Suppl.\  {\bf 192}, 17 (2011)
  [\href{https://arxiv.org/abs/1001.4758}{{\blu{arXiv:1001.4758}}}].
 
  
\bibitem{Pullen:2010zy} 
  A.~R.~Pullen and C.~M.~Hirata,
  JCAP {\bf 1005}, 027 (2010)
    [\href{https://arxiv.org/abs/1003.0673}{{\blu{arXiv:1003.0673}}}].
  
\bibitem{Ade:2013nlj} 
  P.~A.~R.~Ade {\it et al.} [Planck Collaboration],
  Astron.\ Astrophys.\  {\bf 571}, A23 (2014)
    [\href{https://arxiv.org/abs/1303.5083}{{\blu{arXiv:1303.5083}}}].
  
\bibitem{Kim:2013gka} 
  J.~Kim and E.~Komatsu,
  Phys.\ Rev.\ D {\bf 88}, 101301 (2013)
    [\href{https://arxiv.org/abs/1310.1605}{{\blu{arXiv:1310.1605}}}].

  

    
    
    
    
    
    

  
\bibitem{ArkaniHamed:2002sp} 
  N.~Arkani-Hamed, H.~Georgi and M.~D.~Schwartz,
  Annals Phys.\  {\bf 305}, 96 (2003)
   [\href{https://arxiv.org/abs/hep-th/0210184}{{\blu{arXiv:0210184}}}].
  
  
\bibitem{Kehagias:2017cym} 
  A.~Kehagias and A.~Riotto, accepted for publication in JCAP
      [\href{https://arxiv.org/abs/1705.05834}{{\blu{arXiv:1705.05834}}}].

  
\bibitem{Buchbinder:1999be} 
  I.~L.~Buchbinder, V.~A.~Krykhtin and V.~D.~Pershin,
  Phys.\ Lett.\ B {\bf 466}, 216 (1999)
   [\href{https://arxiv.org/abs/hep-th/9908028}{{\blu{arXiv:9908028}}}].
  

\bibitem{Riotto:2008mv} 
  A.~Riotto and M.~S.~Sloth,
  JCAP {\bf 0804}, 030 (2008)
      [\href{https://arxiv.org/abs/0801.1845}{{\blu{arXiv:0801.1845}}}].
  
\bibitem{Burgess:2009bs} 
  C.~P.~Burgess, L.~Leblond, R.~Holman and S.~Shandera,
  JCAP {\bf 1003}, 033 (2010)
      [\href{https://arxiv.org/abs/0912.1608}{{\blu{arXiv:0912.1608}}}].

\bibitem{Seery:2010kh} 
  D.~Seery,
  Class.\ Quant.\ Grav.\  {\bf 27}, 124005 (2010)
    [\href{https://arxiv.org/abs/1005.1649}{{\blu{arXiv:1005.1649}}}].
  
\bibitem{Creminelli:2012ed} 
  P.~Creminelli, J.~Norena and M.~Simonovic,
  JCAP {\bf 1207}, 052 (2012)
     [\href{https://arxiv.org/abs/1203.4595}{{\blu{arXiv:1203.4595}}}].
 



\end{thebibliography}
\end{document}